\documentclass[journal]{IEEEtran}
\usepackage{graphicx}
\usepackage{amsfonts}
\usepackage{amsmath}
\usepackage[table,xcdraw]{xcolor}
\usepackage{graphicx}

\usepackage{amsmath}
\usepackage{array}
\usepackage{caption}
\usepackage{floatrow}   
\usepackage{subcaption}
\usepackage{graphicx,xcolor} 
\usepackage{breqn}
\usepackage[framemethod=tikz]{mdframed}
\usepackage[tableposition=top]{caption}
\usepackage{float}
\floatstyle{plaintop}
\restylefloat{table}
\usepackage{arydshln}
\usepackage{hyperref}
\usepackage{tabularx}
\usepackage{ifpdf}
\usepackage{booktabs}
\usepackage{booktabs}
\usepackage{graphicx}
\usepackage{adjustbox}
\usepackage{cite}
\usepackage[numbers,sort]{natbib}
\usepackage{csquotes}
\definecolor{Orange}{rgb}{1,0.5,0}

\ifCLASSINFOpdf

\else

\fi

\begin{document}
	
	\title{Robust Scatterer Number Density Segmentation of Ultrasound Images}

	\author{Ali K. Z. Tehrani, Ivan M. Rosado-Mendez, and Hassan Rivaz
		\thanks{A. K. Z. Tehrani and H. Rivaz are with the Department
			of Electrical and Computer Engineering, Concordia University, Canada.	
			Ivan M. Rosado-Mendez is with the Department of Medical Physics, University of Wisconsin, United States.
			e-mail: A\_Kafaei@encs.concordia.ca, rosadomendez@wisc.edu and 
			hrivaz@ece.concordia.ca}%
		\thanks{}}

	\markboth{Journal of \LaTeX\ Class Files,~Vol.~14, No.~8, August~2015}%
	{Shell \MakeLowercase{\textit{et al.}}: Bare Demo of IEEEtran.cls for IEEE Journals}
	
	\maketitle

	\begin{abstract}
		Quantitative UltraSound (QUS) aims to reveal information about the tissue microstructure using backscattered echo signals from clinical scanners. Among different QUS parameters, scatterer number density is an important property that can affect estimation of other QUS parameters. Scatterer number density can be classified into high or low scatterer densities. If there are more than 10 scatterers inside the resolution cell, the envelope data is considered as Fully Developed Speckle (FDS) and otherwise, as Under Developed Speckle (UDS). In conventional methods, the envelope data is divided into small overlapping windows (a strategy here we refer to as patching), and statistical parameters such as SNR and skewness are employed to classify each patch of envelope data. However, these parameters are system dependent meaning that their distribution can change by the imaging settings and patch size. Therefore, reference phantoms which have known scatterer number density are imaged with the same imaging settings to mitigate system dependency. In this paper, we aim to segment regions of ultrasound data without any patching. A large dataset is generated which has different shapes of scatterer number density and mean scatterer amplitude using a fast simulation method. We employ a convolutional neural network (CNN) for the segmentation task and investigate the effect of domain shift when the network is tested on different datasets with different imaging settings. Nakagami parametric image is employed for the multi-task learning to improve the performance. Furthermore, inspired by the reference phantom methods in QUS, A domain adaptation stage is proposed which requires only two frames of data from FDS and UDS classes. We evaluate our method for different experimental phantoms and \textit{in vivo} data. 
	\end{abstract}
	

	\IEEEpeerreviewmaketitle

	\section{Introduction}
	Ultrasound imaging is being increasingly used in different clinical applications by providing a gray-scale image that qualitatively shows tissue anatomy. The microstructures within the tissue which are smaller than the wavelength of the ultrasound waves scatter the wave.  
	Quantitative UltraSound (QUS) aims to use backscattererd signal to reveal information about the scatterers, which is also highly correlated with tissue characteristics \cite{ oelze2016review,rouyer2016vivo,nam2011simultaneous}.
	
	Spectral and speckle statistics analysis of backscattered signal has been used in QUS. Spectral-based methods are employed to obtain frequency-dependent quantitative parameters such as the backscattered coefficient, the Effective Scatterer Diameter (ESD) and Effective Acoustic Concentration (EAC) \cite{jafarpisheh2020regularized,Vajihi2018,nam2011simultaneous}. In speckle statistics analysis, the envelope of Radio Frequency (RF) data has been extensively used. Envelope-based methods have been employed to estimate QUS parameters such as scatterer number density and mean scatterer amplitude \cite{destrempes2013estimation}. First-order statistics of envelope data is mostly used in the envelope-based methods \cite{Rosado2016,rivaz2006p3e}.   
	
	The scatterer number density is an important property of the tissue which is defined as the number of scatterers per resolution cell (an ellipsoidal volume defined by - 6 dB  point  of  the  beam  profile \cite{wagner1983statistics}). This quantitative parameter can affect the estimation of other parameters such as ESD; therefore, it is important to characterize the tissue based on this parameter. If the number of scatterers per resolution cell is high (more than 10 per resolution cell), the envelope data is considered as Fully Developed Speckle (FDS), and otherwise, as Under Developed Speckle (UDS) \cite{dutt1995speckle,rivaz2006p3e,Rosado2016}. 
	
	Statistical parameters are mostly used to classify the scatterer number density. If the region is FDS, the distribution of the envelope data can be modeled as the Rayleigh distribution \cite{wagner1983statistics}. When there are a few scatterers within the resolution cell, the Rayleigh distribution fails to model the envelope data, and other more general distributions such as the K-distribution,  the homodyned K-distribution \cite{destrempes2013estimation}, and the Nakagami distribution \cite{shankar2001classification,ho2012using} have been proposed to model the envelope data. Among these distributions, the Nakagami distribution has attracted attention of many researchers due to its low-complexity and ability to perform closed-form parameter estimation, and has been used to characterize many different tissue types \cite{ho2012using,tsui2008performance}.

	In conventional envelope-based methods, the envelope data is divided into small overlapping windows (a strategy here we refer to as patching), and an inverse problem approach is used to estimate the scatterer number density and other QUS parameters \cite{Rosado2016,destrempes2013estimation}. The size of the patch should be large enough to contain adequate independent samples. Large patches reduce the spatial resolution and cannot identify small regions having different QUS parameters. The size of the patch is an important hyper-parameter that the user should optimize.  
	
	Convolutional Neural Networks (CNN) have been widely used in ultrasound applications including elastography \cite{tehrani2020displacement,tehrani2020semi,tehrani2021mpwc,delaunay2021unsupervised}, segmentation \cite{amiri2020fine,karimi2019accurate} and phase aberration correction \cite{sharifzadeh2020phase,jeon2020deep}. However, they have been used rarely for QUS due to requirement of large training data and the system specific nature of quantitative applications. Recently, a patch-based CNN was developed by our group to classify regions in the ultrasound image according to their scatterer number density \cite{tehrani2021ultrasound}. This prior work was one of the first reports on the use of CNN on QUS, particularly for scatterer number density classification and resulted in promising results (AUC as high as 0.975 in tissue mimicking phantoms). However, the patch-based nature of this work causes limitations especially when applying the method to real world scenarios where there are regions with irregular shapes with heterogeneous composition. The present work significantly advances QUS methods based on speckle statistics classification by avoiding the use of patches.  
	
	In this paper, we segment the scatterer number density of the whole ultrasound image using a fully convolutional neural network. A simple method is introduced to generate a large dataset having different scatterer number density and mean scatterer amplitude shapes. The generated dataset enables us to classify samples of ultrasound envelope data without patching requirement. Furthermore, we uniformly sample the imaging parameters from a wide range to make the trained network robust to change of imaging parameters. We then use the Nakagami parameter in a multi-task manner to reduce over-fitting of the network. The idea of the reference-based methods \cite{yao1990backscatter,Vajihi2018,Rosado2016,jafarpisheh2020analytic} is also adapted for CNNs to further boost the performance. We validate our method using simulation data, experimental phantoms and \textit{in vivo data}. Our contributions are summarized as follows:
	\begin{itemize}
		\item The scatterer number density is estimated for the whole image instead of estimating for each individual patch. To the best of our knowledge, this is the first approach without patching for segmentation based on scatterer number density.
		\item The proposed method is evaluated for a variety of different computational and physical phantoms and \textit{in vivo} breast ultrasound, imaged with different ultrasound scanners and transducers.
		\item Nakagami parametric imaging is employed as multi-task learning to improve the performance of network.
		\item The reference phantom idea of conventional methods is adopted to CNN for domain transformation to further enhance the performance using only a small amount of phantom data. 
		\item The proposed method is compared with a previously developed reference-based method and a patch-based CNN recently proposed by our group.   
		\item A fast and simple ultrasound image generation is employed which enables us to generate thousands of images with diverse imaging and scatterer properties for training our segmentation network. The generated dataset contains different scatterer number density and mean scatterer amplitude. 
	\end{itemize}

	\section{Methods}

	\subsection{Background}
	\label{sec:RW}
	
	\label{sec:pagestyle}
	
	\subsubsection{Conventional methods to evaluate the scatterer number density}
	\label{sec:ref}
	Rosado-Mendez \textit{et al.} used a window around the region of the interest (here, we call it a patch) and employed echo amplitude signal-to-noise ratio, SNR, defined in Eq \ref{eq:snr}, the generalize spectrum and the first order statistics of the phase information to classify ultrasound patches into different groups of low-scatterer number density, diffuse scattering and coherent scattering \cite{Rosado2016}. 
	\begin{equation}
		\label{eq:snr}
		\mathrm{SNR}= \frac{(mean[A])}{\sqrt{Var[A]}}
	\end{equation}
	In Eq \ref{eq:snr}, $A$ denotes envelope. The SNR increases with scatterer number density. If the patch is FDS, this value would be close to 1.91 \cite{wagner1983statistics}. A reference phantom with high scatterer number density was used to account for spatial variations of the resolution cell size due to diffraction effects. A patch of ultrasound envelope data was classified as FDS if the SNR value was close to the SNR value of the reference phantom patch from the same depth. If the SNR value was lower than the reference one, the patch was considered as UDS and if it was higher, it was considered as non-resolved periodicity. Here, we assumed that if the absolute difference was lower than 3\% (it was set empirically) of the reference SNR, the region is considered as FDS. Our method is compared with this algorithm (here it is called \enquote{Reference Method}) for experimental phantom evaluations. 
	\subsubsection{Deep learning methods}
	In \cite{tehrani2020pilot}, we proposed a CNN to classify scatterer number density of a small patch of envelope data. The network was fed with envelope data and the spectrum of RF data, and was compared with a Multi-Layer Perceptron (MLP) classifier, which used two statistical parameters (signal to noise ratio and skewness) as inputs. The method achieved a segmentation  accuracy of 92.2\% for simulation test data with a patch size of $1.5mm\times 1.5mm$. However, it was limited to a single imaging setting and was also a patch-based method that required to apply the network many times over multiple patches to generate the scatterer number density map of the whole image. Furthermore, the variation of mean scatterer amplitude was not considered. 
	
	In \cite{amiri2020segmentation}, we used a U-Net to segment scatterer number density of ultrasound images. Simulation data having inclusions with different densities were generated. The results showed that the network was able to segment the simulation test data with 1 and 10 scatterers per resolution cell with a precision of 99.2\% and 67.5\% and sensitivity of 98.8\% and 79.7\%, respectively, provided that there was a noticeable difference in intensity of the corresponding regions. However, the change of amplitude due to mean scatterer amplitude was not considered.

	Zhang \textit{et al.} \cite{zhang2020deep} recently used a U-Net to estimate the pixel-wise mean scattering intensity. They generated a dataset with random shapes and considered different values of mean scattering intensity for each region. They assumed that all areas were FDS; hence, the network was able to associate different values of intensity to the mean scattering intensity of the scatterer distribution. However, the FDS assumption does not hold for many organs and limits the generalization to FDS tissues.
	
	In our recent work \cite{tehrani2021ultrasound}, we used state-of-art CNN architectures as well as patch statistics to classify scatterer number density of patches. We simulated a training dataset with a fixed imaging parameters using Field II \cite{jensen1996field} which is available online at \url{code.sonography.ai}. CNNs which employed envelope echo signals outperformed machine learning methods such as Support Vector Machine (SVM), Random Forest and Multi-Layer Perceptron which only used patch statistics. Fusion and Multi-Task Learning (MTL) were also utilized to combine the information of the statistics and textures. MTL was shown to be an appropirate choice for unseen experimental phantom data. This method is used for comparison and is labeled as \enquote{Patch-based CNN}.

	\subsection{Datasets and Data Generation}
	
	\subsubsection{Data Generation}
	\label{sec:data_gen}
	A large and diverse dataset is required to train a fully convolutional neural network. In fact, one of the reasons that patch based methods for scatterer classification were developed was to reduce data requirements of CNNs. 
	
	In this section, a simple but effective data generation scheme is introduced. Ultrasound simulation tools such as Field II \cite{jensen1996field} can be used to generate ultrasound simulation data. These methods are computationally expensive but have been used to generate medium size datasets such as the dataset proposed in \cite{tehrani2020displacement,tehrani2021ultrasound}. Field II takes several minutes on a typical machine to generate an image. A common trend to generate medium size datasets is to use clusters but it is still infeasible to generate very large datasets (more than 10,000 images). 
	
	We employ a simple method to generate a large number of images in a short amount of time. This data generation method shares some similarities with the method used in \cite{zhang2020deep}, in that it considers variations in mean scatterer intensity. The main difference of the method proposed here is that it also considers variations in scatterer number density. This makes the data set more realistic and suitable for scatterer number density segmentation. It should be mentioned that unlike Field II simulations, which are scatterer-based and defined in arbitrary coordinates, the approach here (and in \cite{zhang2020deep}) is grid-based, where each discrete grid position is assigned scattering properties.

	Assuming weak scattering (using the first order Born approximation), the interaction of scatterers with ultrasound waves can be modeled by a 2D linear time varying convolution \cite{jensen1993deconvolution,burger2012real,gao2009fast}.    
	\begin{equation}
		r[a,l]=g[a,l]\ast h[a,l]+\eta [a,l]
	\end{equation}
	where $g[a,l]$ is the scatterer echogenicity map, $h[a,l]$ is the spatially varying Point Spread Function (PSF) at axial and lateral positions $a$, $l$, and $\eta$ indicates additive white Gaussian noise. The scatterer echogenicity in a specified location is a Gaussian random variable sampled from a Bernoulli distribution which can be written as \cite{zhang2020deep}:
	\begin{equation}
			\label{eq:g}
			g(a,l)=K(a,l)\times A(a,l)
	\end{equation}
where $K(a,l)$ denotes a sample at the axial and lateral location \textit{a} and \textit{l}, respectively, from a Bernoulli distribution with value of 1 with the probability of $p$ (probability of presence of a scatterer) and 0 with the probability of $1-p$. \textit{A} denotes the amplitude of the scatterers and it is sampled from $N(\mu_s,\sigma^2)$ where \textit{N} denotes normal distribution with variance of $\sigma^2$ and mean of $\mu_s$ which corresponds to the mean scatterer amplitude.

	Equation \ref{eq:g} can be used for a phantom having fixed values of mean scatterer amplitude ($\mu_s$) and scatterer number density. In order to incorporate different shapes of scatterer number densities and mean scattering intensity, we define a scatterer number density binary mask ($SC$) and a mean scatterer amplitude binary mask ($MS$) to control $K(a,l)$ and $\mu_s$ as the following:
	\begin{equation}
		\label{eq:MS}
		\begin{aligned}
			K(a,l) =\begin{Bmatrix}
				K_{1}(a,l) & SC=0\\ 
				K_{2}(a,l) & SC=1
			\end{Bmatrix}
			\\
			\mu_{s}(a,l) =\begin{Bmatrix}
				\mu_{s1}(a,l) & MS=0\\ 
				\mu_{s2}(a,l) & MS=1
			\end{Bmatrix}
		\end{aligned}
	\end{equation}
	where $K_{1}(a,l)$ and $K_{2}(a,l)$ are the Bernoulli distributions associated to the different values of scatterer number density binary mask. $\mu_{s1}(a,l)$ and $\mu_{s2}(a,l)$ are different values of mean scatterer amplitude assigned to the different values of mean scatterer amplitude mask. Inserting Eq \ref{eq:MS} into Eq \ref{eq:g} leads to:
	\begin{equation}
			\
			\label{eq:g2}
			\begin{aligned}
				\resizebox{0.45\textwidth}{!}{$
					g=\begin{Bmatrix}
						\sum_{a}^{}\sum_{l}^{}K_{1}(a,l)\times A(a,l)& SC=0,MS=0\\ 
						\sum_{a}^{}\sum_{l}^{}K_{1}(a,l)\times A(a,l)& SC=0,MS=1\\ 
						\sum_{a}^{}\sum_{l}^{}K_{2}(a,l)\times A(a,l)& SC=1,MS=0\\
						\sum_{a}^{}\sum_{l}^{}K_{2}(a,l)\times A(a,l) &SC=1,MS=1\\
					\end{Bmatrix}$}
		\end{aligned}
	\end{equation}
	The Eq \ref{eq:g2} can be extended to more than 4 states by using non-binary $SC$ and $MS$ masks.

	We consider the simplifying assumption that PSF ($h[a,l]$) is constant throughout the image. Therefore, the time varying convolution is converted to a time invariant one. The PSF can be modelled by a 2D Gaussian function modulated by a cosine function in axial direction \cite{gao2009fast,burger2012real}.
	\begin{equation}
		h[a,l]=e^{-\frac{1}{2}(\frac{a^2}{{\sigma_{a}}^{2}}+ \frac{l^2}{{\sigma_{l}}^{2}})} \times cos(2\pi f_{c}a)  
	\end{equation}
	where $f_{c}$, ${\sigma_{a}}^{2}$ and ${\sigma_{l}}^{2}$ denote the center frequency, axial and lateral width of the Gaussian profile of the PSF, respectively.
	
\begin{figure}[!t]
		\centering
		\includegraphics[width=0.95\textwidth]{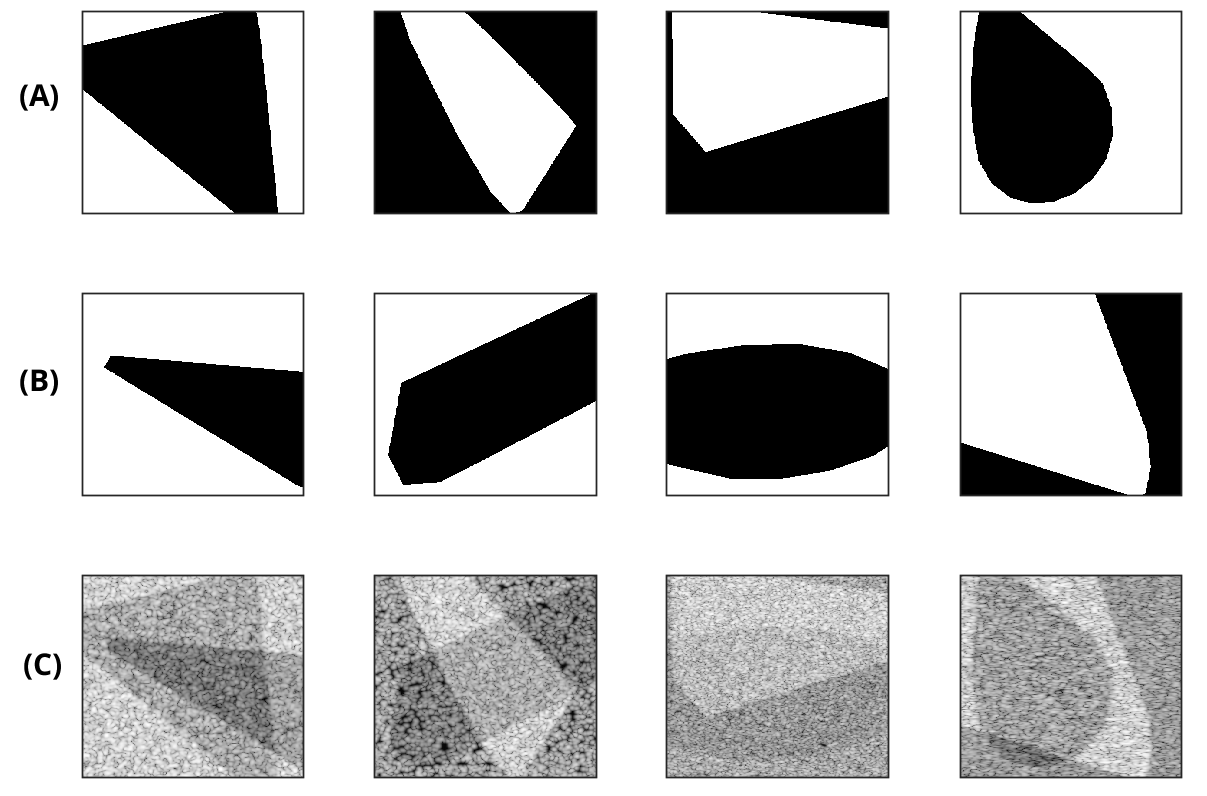}
		\centering
		\caption{Four examples of the generated dataset. Scatterer number density mask (A), mean scatterer amplitude mask ($\mu_s$) (B) and obtained B-mode image (C). }
		\label{fig:generated_data}
	\end{figure}

	We generated 7000 random binary shapes and assigned different scatterer number densities and mean scatterer amplitude to each region. The imaging parameters $\sigma_s$, $\sigma_l$, $f_c$ and speed of sound ($v$) are also sampled from a uniform distribution. The imaging parameters and their ranges are specified in Table \ref{tab:tab:gen}. We generated 15000 images for training and validation using 6500 random binary masks, and an additional 500 test images using 500 binary masks. The test binary masks are not used in training data generation to avoid data leakage. Some examples of the generated dataset are illustrated in Fig. \ref{fig:generated_data}. It can be seen from the figure that the intensity of the output image largely depends on both scatterer number density and mean scatterer amplitude; therefore, both of them must be taken into account when dealing with ultrasound images. 
	
\begin{figure*}[!t]
		\centering
		\includegraphics[width=0.8\textwidth]{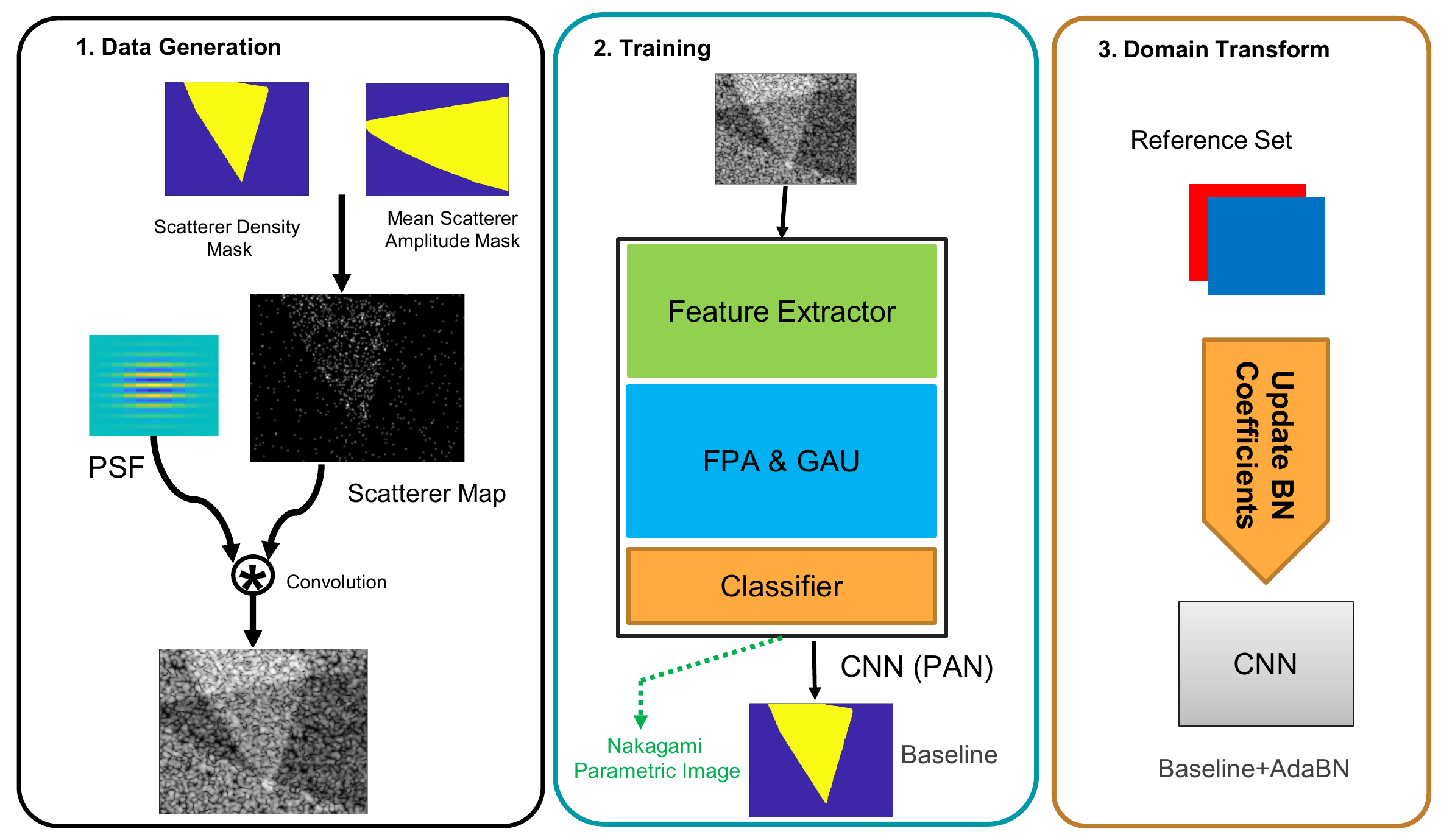}
		\centering
		\caption{Overview of the framework used in the paper.}
		\label{fig:arch}
	\end{figure*}

	In \cite{zhang2020deep}, the ultrasound images were assumed to be FDS; therefore, they associated the intensity to the mean scatterer amplitude. In contrast to our recent work \cite{amiri2020segmentation}, it was assumed that all ultrasound images had the same mean scatterer amplitude; therefore, the images could be segmented by considering the intensity. In contrast to the previous works, the present method considers that both mean scatterer amplitude and scatterer number density can vary independently. Figure \ref{fig:arch} part 1 depicts the training data generation step. The test set of this dataset will be available online after acceptance of this manuscript at \url{code.sonography.ai} similar to our previous works~\cite{tehrani2020displacement,tehrani2021ultrasound}.          
	
	\begin{table}[]
		\caption{Parameters of image generation and their ranges.}
		\label{tab:tab:gen}
		\resizebox{\textwidth}{!}{%
			\begin{tabular}{@{}cc@{}}
				\toprule
				\textbf{Parameter}                                      & \textbf{Range}    \\ \midrule
				Scatterer Number Density (UDS)                                 & 1-2               \\
				\rowcolor[HTML]{EFEFEF} 
				Scatterer Number Density (FDS)                                 & 11-16             \\
				Mean Scatterer Amplitude ($\mu_s$)                 & 0.3-1.3           \\
				\rowcolor[HTML]{EFEFEF} 
				Standard Deviation of Scattering Amplitude ($\sigma_s$) & 0.03 (fixed)              \\
				Center Frequency ($f_c$)                                & 4 - 7 MHz         \\
				\rowcolor[HTML]{EFEFEF} 
				Sampling Frequency ($f_s$)                              & 60 - 100 MHz      \\
				Speed of Sound ($v$)                                    & 1510 - 1560 $m/s$ \\
				\rowcolor[HTML]{EFEFEF} 
				F Number                                                & 1.5 - 2.5         \\
				Number of Excitation Pulses                             & [3-5]           \\
				\rowcolor[HTML]{EFEFEF} 
				Standard Deviation of PSF in axial ($\sigma_a$)         & 0.1 - 0.3  $mm$       \\
				Standard Deviation of PSF in lateral ($\sigma_l$)       & 0.13 - 0.4 $mm$       \\ \bottomrule
			\end{tabular}%
		}
	\end{table}

	\subsubsection{Experimental Phantoms}
	Three homogeneous phantoms, A, B and C are employed for the evaluation. They had a size of 15cm$\times$ 5cm$\times$ 15cm, made from a mixture of agarose gel media and glass beads as scattering agents. The phantoms have different scatterer number density and mean scattering amplitudes, more details are given in Table \ref{table:exp_phantom}.  For more information about these phantoms refer to \cite{rosado2014advanced}. An Acuson S2000 scanner (Siemens Medical Solutions, Malvern, PA) with an 18L6 transducer having the center frequency of 8.89 MHz was used to image the phantoms and RF data was acquired using Axius Direct Ultrasound Research Interface \cite{brunke2007ultrasound}. We used the correlation method to compute the resolution cell size \cite{rivaz20079c}. The axial and lateral correlation cell size were 0.26 $mm$ and 0.49 $mm$ at the focal point, respectively.
	
	\begin{table}[!t]
		\caption{Characteristics of the homogeneous phantoms.}
		\label{table:exp_phantom}
		\resizebox{\textwidth}{!}{%
			\begin{tabular}{@{}cccc@{}}
				\toprule
				Parameter                                   & Phantom A & \multicolumn{1}{l}{Phantom B} & \multicolumn{1}{l}{Phantom C} \\ \midrule
				Diameter of Random Scatterers ($\mu m$)     & 5-40      & 75-90                         & 126-151                       \\
				Scatterer Concentration per $mm^3$          & 236       & 9                             & 3                             \\
				Scatterer Concentration per resolution cell & 8.50-67   & 0.32-2.55                     & 0.11-0.85                     \\ \bottomrule
			\end{tabular}%
		}
	\end{table}
	
	\subsubsection{Phantom with Inclusions (Phantom D)}
	This phantom was imaged by a Verasonics Vantage 128
	system (Verasonics, Kirkland, WA) using a L11-5v transducer
	operated at 8 MHz. This phantom has three inclusions with different echogenicities and scatterer densities \cite{jafarpisheh2020analytic}. We computed the resolution cell size using correlation method \cite{rivaz20079c}, which was 0.149 $mm$ and 0.237 $mm$ at the focal point in the axial and lateral direction, respectively.
	
	\subsubsection{CIRS phantom (Phantom E)}
	A multi-purpose, multi-tissue CIRS phantom (model 040GSE, Norfolk, VA) was also used in evaluating the performance of the proposed method. It contains inclusions with no scatterers and different scatterer number densities. Data was collected with an E-CUBE 12 Alpinion machine using L3-12H transducer. The center and sample frequencies were 8.5 MHz and 40 MHz, respectively.
	
The background has high scatterer number density (FDS) and there are inclusions with different scatterer number densities. We selected this phantom to evaluate the proposed method for the cases where the scatterer number density was between the lowest (no scatterer) and the highest (FDS).
	
	The phantom was also imaged using Verasonics Vantage 256 system (Verasonics, Kirkland, WA) using a L11-5v transducer. The results using this scanner are given in the Supplementary Material. No reference phantom is available for this phantom; therefore, the proposed method is only compared with the patch-based CNN.
	
	\subsubsection{\textit{In vivo} data}
	We used breast ultrasound images recorded by a Siemens Sonoline Elegra System (Issaquah, WA) with the sampling frequency of 36 MHz, center frequency of 7.5 MHz and a lateral beam spacing of 200 µm. I-Q echo data were recorded in a file on the imaging system when data acquisition was stopped (frozen on the imaging system). The IQ data were converted to RF echo data offline using the known demodulation frequency of the imaging system. More information about this dataset and the recording procedure is provided in \cite{zhu2002modified}.

	\subsection{Nakagami Parametric image}
	The Nakagami distribution is a flexible tool to model different scatterer patterns based on envelope amplitudes. It can be defined as \cite{shankar2001classification}:
	\begin{equation}
		f(A,m,\Omega )=\frac{2m^{m}}{\Gamma(m)\Omega^m}A^{(2m-1)}\times e^{-\frac{m}{\Omega}A^2}
	\end{equation}
	where $A$ denotes the envelope amplitude, $m$ represents the shape parameter, $\Omega$ is the scale parameter and $\Gamma$ denotes the Gamma function. $m$ is found to be correlated with scatterer number density, values close to 1 reflect high scatterer number density, and low values represent low scatterer number density. $m$ can be estimated by the maximum likelihood method \cite{shankar2001classification,Rosado2016}. The Nakagami parameter is also machine dependent meaning that machine settings can change the value of $m$ for the same tissue \cite{tsui2017effect}.
	
\begin{figure}[!t]
		\centering
		\includegraphics[width=0.99\textwidth]{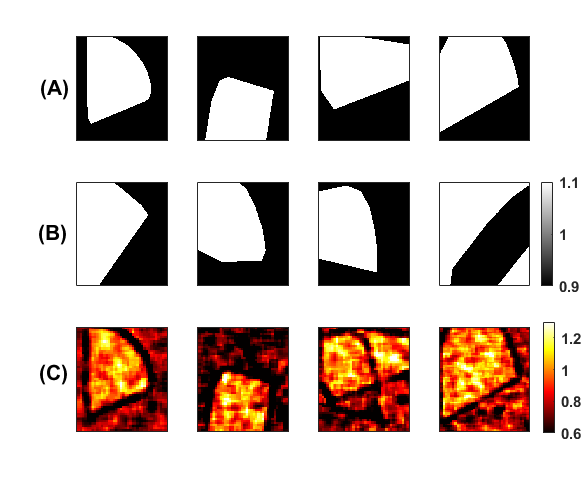}
		\centering
		\caption{Four examples of the generated dataset. Scatterer number density mask (A), mean scatterer amplitude mask ($\mu_s$) (B) and obtained Nakagami parametric image $m$ (C). The mean amplitude values in these 4 samples are either 0.9 or 1.1. }
		\label{fig:m}
	\end{figure}

	In order to obtain the parametric image of $m$, patches of envelope data with overlaps are extracted and the $m$ parameter is estimated using maximum likelihood estimator \cite{Rosado2016,tsui2008performance}. The patches must be large enough to provide statistically reliable estimates of $m$. But very large patches reduce the spatial resolution of the Nakagami parametric image and might result in loss of information especially for small targets. We ensured that the window for estimation of the Nakagami parameter is at least 8 times larger than the resolution cell size. Some examples of the obtained parametric images are shown in Fig. \ref{fig:m}, which shows that the Nakagami parameter is mostly sensitive to the c6hanges of scatterer number density, whereas brightness changes in the envelop images have little effects on Nakagami parameter. We used Nakagami parametric images in two different fashions; as the input of the network and as an auxiliary output for multi-task learning \cite{zhang2018overview}.         
	\subsection{Network Architecture and Training}
	Semantic segmentation networks are able to segment more than 200 different classes within an image. Among different networks, the Pyramid Attention Network (PAN) \cite{li2018pyramid} has shown its superior performance over other well-known networks such as PSPNet \cite{zhao2017pyramid} and DeepLab v2 \cite{chen2017deeplab}. We used this network for our segmentation task. The network is composed of three main blocks.

	1) Feature extractor: this module is employed to extract features from raw input images. ResNet50 \cite{he2016deep} is utilized for the feature extraction. To avoid loss of information, features from different levels of ResNet50 are used to keep both spatial and semantic information.

	2) Feature Pyramid Attention (FPA): This module is an attention module and it is used to have a precise pixel-level attention for high level features \cite{zhao2017pyramid}.

	3) Global Attention Upsample (GAU): This module provides channel-wise attention mechanism to emphasize more important channels of low-level features using high-level features.
	
	This network was employed to obtain scatterer number density segmentation with 1/4 of the original input resolution. It should be noted that the aim of this paper is to investigate the performance of a state of the art CNN in scatterer number density segmentation, and not to compare different networks for this task.

	The loss function for the segmentation task is the summation of binary cross entropy and the Dice loss. For the MTL network (with auxilary output), the loss is added by smooth L1 loss of the Nakagami parametric image. The loss for this network can be written as:
	\begin{equation} 
		\begin{gathered}
			loss = BCE(\widehat{Ds},Ds_{gt}) + Dice(\widehat{Ds},Ds_{gt})\\ + \beta |\widehat{m}-m_{gt}|_1,
		\end{gathered} 
	\end{equation}
	where BCE, Dice and $|.|_1$ denote binary cross entropy, Dice loss and smooth L1 norm, respectively. $\widehat{Ds}$, $Ds_{gt}$, $\widehat{m}$ and $m_{gt}$ represent predicted, ground truth scatterer number density and predicted and ground truth Nakagami parametric image, respectively. $\beta$ is the weight associated to the auxiliary loss (MTL) of the estimated Nakagami parametric image which is set to 0.1 to have lower weight than the main task. The auxiliary loss can be viewed as a regularizer that avoids over-fitting to the training data. The auxiliary task should be related to the main task \cite{arik2017deep}; therefore, we selected the Nakagami parameter since it is highly correlated with the scatterer number density.

	The networks were trained using the Adam optimizer for 20 epochs and the weights with the best validation results were used for evaluation. The learning rate was set to 1e-5 for the first 10 epochs and then reduced to 1e-6 for the last 10 epochs. It should be noted that we used feature extraction block pre-trained on ImageNet to speed up the training. The top performing network weights will be available online after acceptance of the manuscript at \url{code.sonography.ai}.

	\begin{figure}[!t]
		\centering
		\includegraphics[width=0.99\textwidth]{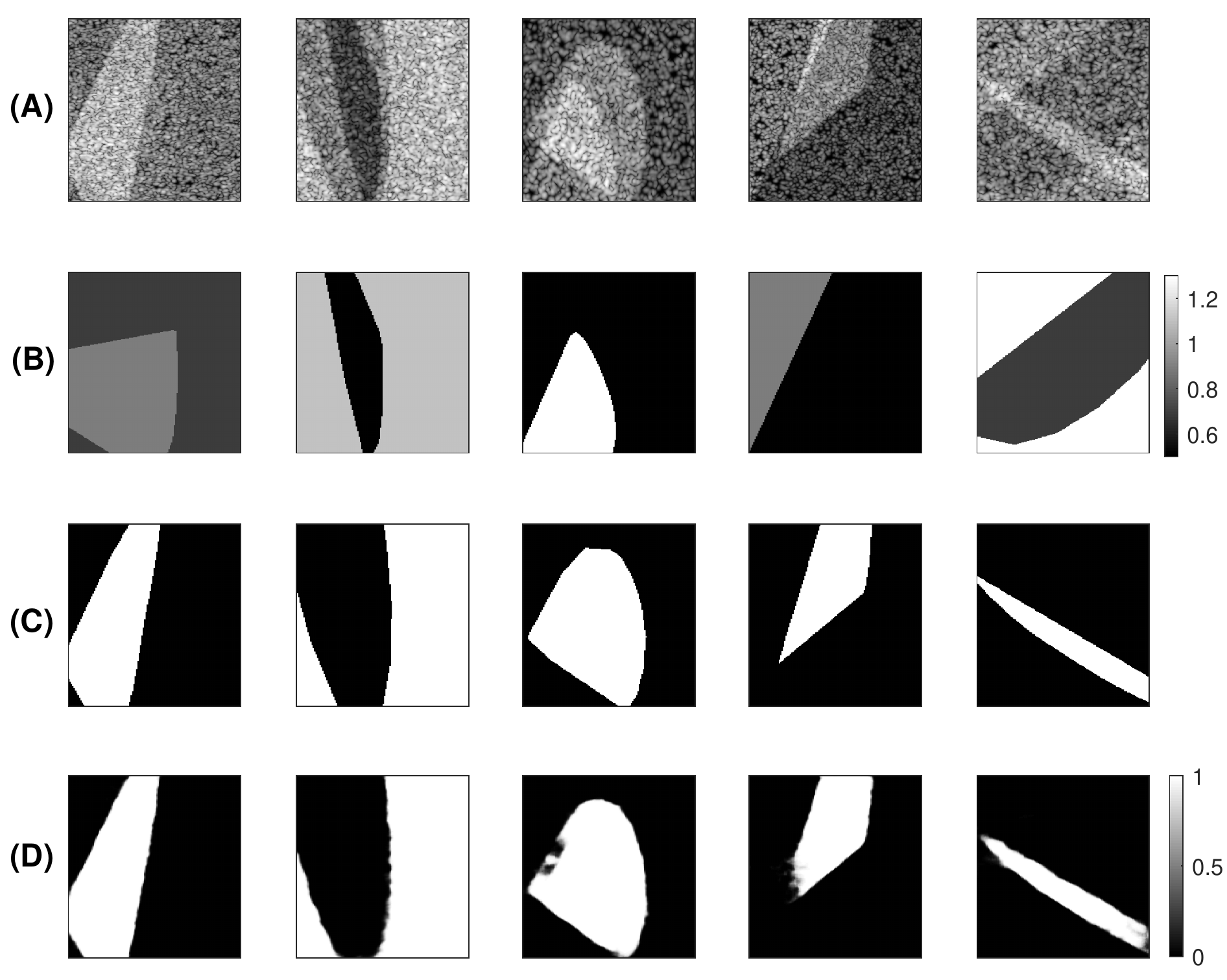}
		\centering
		\caption{Four examples of the generated dataset with different imaging parameters. B-mode images (A), mean scatterer amplitude mask ($\mu_s$) (B), scatterer number density mask (C) and the output of Baseline (D).}
		\label{fig:sim}
	\end{figure}

	\subsection{Batch Normalization and Adaptive Batch Normalization}
	\label{sec:adabn}
	Batch Normalization (BN) has been found to be very useful in deep learning networks. It can speed up the training and remove covariance shift \cite{ioffe2015batch}. Having the input of the BN layer $X\in \mathbb{R}^{n\times c \times h \times w}$, where $n$ is the number of data in a mini-batch, $c$, $h$ and $w$ denote the number of channels, feature height and width, respectively. The BN layer performs the following operation on $X$:
	\begin{equation}
		\label{eq:batch}
		\begin{gathered}
			\widehat{x}_j=\frac{x_j-E[X_j]}{\sqrt{Var[X_j]}},\\
			y_j = \gamma_j \widehat{x}_j + \beta_j,
		\end{gathered}
	\end{equation}
	where $j$ is index of channel, $x_j$ and $y_j$ denote input/output of the BN layer for channel $j$ of one data sample. $\gamma_j$ and $\beta_j$ are learnable parameters that are optimized during the training. The mean ($E[X_j]$) and variance ($Var[X_j]$) of the mini-batch are computed in the training. During the test time, the computed coefficients in the training are used for normalization. By applying Eq \ref{eq:batch}, the distribution of all mini-batches remains the same. Considering a binary classification problem, if the size of mini-batch is too small that data belonging to only one class appears inside the mini batch, the obtained statistics would be biased toward that class which degrades performance. To mitigate this problem, other normalization techniques such as Group Normalization has been proposed for small mini-batch sizes \cite{wu2018group}.
	
	Another aspect of BN is its impact on domain shift. The weights of the networks mostly preserve information about the classes, whereas, the BN coefficients contain information about the domain of the training data \cite{li2018adaptive}. Motivated by this, the Adaptive Batch Normalization (AdaBN) has been proposed for domain adaptation. The basic idea of AdaBN is that the coefficients of BN obtained during training are not suitable for the test data if there is a domain shift.

	In AdaBN, the coefficients of the BN layers are updated using data belonging to the test domain \cite{li2018adaptive}. The main difference between AdaBN and transfer learning (fine-tuning) is that the weights of the network are not altered; therefore, only a small amount of data belonging to the test domain is required to update the BN coefficient. We call this data \enquote{reference set} since it acts very similar to the reference phantoms used in QUS. The reference set must have a balanced amount of data from each class to avoid a biased mean and variance. The reference set can be as small as two frames of data for binary classification task. In fact, for domain transformation only parameters of Eq \ref{eq:batch} are required to be learned. In our experiment, two frames from phantom A (class 1) and phantom B (class 0) were employed. This method of domain transformation requires very small amount of data and does not leads to over-fitting since only a few parameters are learned by back propagation. AdaBN is well-suited for our task since the effect of imaging parameters can be reduced using domain transformation techniques. It can be used to transfer the baseline to any machine setting by only small amount of data. 
	
	\begin{table*}[]
		\caption{The mean and standard deviation of the simulation results of scatterer number density segmentation. Baseline-nm, Baseline-ni and Baseline are from PAN network.}
		\label{tab:sim}
		\adjustbox{width=\textwidth}{
			\begin{tabular}{ccccccccccccc}
				& \multicolumn{3}{c}{\textbf{IOU}} & \multicolumn{3}{c}{\textbf{Accuracy}} & \multicolumn{3}{c}{\textbf{Sensitivity}} & \multicolumn{3}{c}{\textbf{Precision}} \\ \cline{2-13} 
				& top 100\%  & top 10\%  & top 5\% & top 100\%    & top 10\%   & top 5\%   & top 100\%     & top 10\%    & top 5\%    & top 100\%    & top 10\%    & top 5\%   \\ \hline \\
				U-Net        &0.950$\pm$0.038  &0.786$\pm$0.068  &0.711$\pm$0.076  &0.973$\pm$0.023  &0.858$\pm$0.050  &0.796$\pm$0.060  &0.973$\pm$0.020  &0.823$\pm$0.042  &0.742$\pm$0.050  &0.976$\pm$0.031  &0.905$\pm$0.058  &0.878$\pm$0.066  \\  \hline \\
				Baseline-nm &0.976$\pm$0.023  &0.922$\pm$0.037  &0.898$\pm$0.040  & 0.989$\pm$0.015 &0.958$\pm$0.032  & 0.938$\pm$0.035 & 0.987$\pm$0.019 & 0.945$\pm$0.037 & 0.922$\pm$0.041 & 0.988$\pm$0.012 & 0.959$\pm$0.017 & 0.948$\pm$0.017 \\ \\
				Baseline-ni        &0.967$\pm$0.041  &0.875$\pm$0.080  &0.829$\pm$0.092  &0.984$\pm$0.022  &0.931$\pm$0.044  & 0.898$\pm$0.048 &0.986$\pm$0.028  & 0.932$\pm$0.070 &0.895$\pm$0.090  &0.980$\pm$0.031  &0.918$\pm$0.060  &0.890$\pm$0.071  \\ \\
				Baseline      &\textbf{0.981$\pm$0.017}  &\textbf{0.941$\pm$0.023}  &\textbf{0.925$\pm$0.023}  & \textbf{0.991$\pm$0.008} &\textbf{0.972$\pm$0.015}  &\textbf{0.962$\pm$0.015}  &\textbf{0.990$\pm$0.012}  & \textbf{0.963$\pm$0.022} & \textbf{0.948$\pm$0.024} &\textbf{0.990$\pm$0.011}  &\textbf{0.966$\pm$0.016}  &\textbf{0.955$\pm$0.017}  \\ \hline
			\end{tabular}%
		}
	\end{table*}

	\section{Results}
	\subsection{Simulation Results}
		\begin{table}[tp]
		
		\caption{The summary of the networks employed in this paper. The parameters $A$, $Alog(A)$ and $m$ denote envelope, envelope multiplied by log compressed envelope and Nakagami parametric image, respectively.}
		\label{tab:sum}
		\adjustbox{width=\textwidth}{
			\begin{tabular}{@{}cccc@{}}
				\toprule
				Name           & Input            & Multi-task & Training data                                                                                                 \\ \midrule
				Baseline-nm    & $A$ and $Alog(A)$    & No         & Grid-based Simulation                                                                                         \\
				Baseline-ni    & $A$, $Alog(A)$ and $m$ & No         & Grid-based Simulation                                                                                         \\
				Baseline       & $A$ and $Alog(A)$    & yes ($m$)    & Grid-based Simulation                                                                                         \\\midrule
				Baseline-AdaBN & $A$ and $Alog(A)$    & yes ($m$)    & \begin{tabular}[c]{@{}c@{}}Grid-based simulation + \\ updating BN layers using experimental data\end{tabular} \\ \bottomrule
		\end{tabular}}
	\end{table}
	\subsubsection{Ultrasound grid-based simulation}
	The test set of simulation data contains 500 test images obtained by the method explained in section \ref{sec:data_gen}. Intersection Over Union (IOU), accuracy, sensitivity and precision are employed to evaluate scatterer number density segmentation performance. Dice similarity score is excluded since it is highly correlated with IOU. All metrics are reported for the whole test set, 10\% and 5\% of the test set having the worst results. The results are given in Table \ref{tab:sim}. The U-Net architecture is the same as \cite{buda2019association}. \enquote{Baseline-nm} denotes the network (PAN) having $A$ (envelope) and $Alog(A)$ (envelope $\times$ log compressed envelope) as input channels. \enquote{Baseline-ni} denotes the same network having Nakagami parametric image as well as $A$ and $Alog(A)$ as input channels. \enquote{Baseline} is the proposed network with $A$ and $Alog(A)$ as inputs but having Nakagami parametric image as an auxiliary output. A summary of  different methods is presented in Table \ref{tab:sum}.

	According to Table \ref{tab:sim}, PAN performs better than U-Net which is expected due to use of different attention mechanism. Adding Nakagami parametric image as an input channel (Baseline-ni) deteriorates the performance (compared with Baseline-nm); however, adding Nakagami parametric images as an auxiliary output (Baseline) improves the performance in all metrics. The lower performance of the network having Nakagami parametric image as an additional input is ascociated to the fact that the texture of Nakagami parametric image does not have valuable information about the scatterer number density, while, its value matters for prediction of scatterer number density. This result also agrees with our recent work \cite{tehrani2021ultrasound} where we found that MTL performs better than adding the statistics as additional inputs. Some examples of predicted segmentation results by Baseline are depicted in Fig. \ref{fig:sim}. It can be seen that the network can segment the scatterer number density well even in presence of different values of mean scatterer amplitude which results in different intensities. It should be noted that in simulation results, there is no domain shift which leads to high performance in all compared methods. Figure \ref{fig:arch} part 2 depicts the simulation training and test step.
		
	\subsubsection{The Network prediction for different values of scatterer number density}
	\label{sec:sc_range}
	The training data only contained scatterer number densities of 1 to 2 scatterers per resolution cell for UDS and 11 to 16  for FDS. To investigate the performance of the algorithm when used on data with scatterer number densities from 1 to 14 scatterers per resolution cell, 10 phantoms were simulated with  $\left\lbrace 1,2,..,14\right\rbrace $, scatterers per resolution cell. Figure 6 shows the mean value and the 95\% confidence interval (as the shading) of the probability of FDS produced by the network. It is clear that when the scatterer number density is close to 1 or 10, the confidence interval (the shaded area) is small while, for scatterer number densities in the range of 4-7, the confidence intervals are wider. The training data could have higher values of scatterer number density for UDS class. For instance, the UDS class could contain values of 4-6 scatterers per resolution cell. This would result in the reduction of the probability of FDS of the network for those values. However, we simulated the training data which contains the very low scatterer number density for UDS. In this way, the value provides some insights about how far the scatterer number density is from FDS.      
	\begin{figure}[t]
		\centering
		\includegraphics[width=0.78\textwidth]{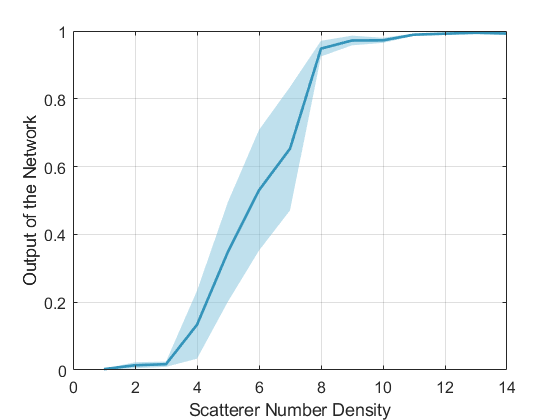}
		\centering
		\caption{The mean and the 95\% confidence interval range of the output of the network for the scatterer number density in the range 1 to 14.}
		\label{fig:range}
	\end{figure} 
	\subsubsection{Field II simulation}
\begin{figure}[t]
	\centering
	\includegraphics[width=0.99\textwidth]{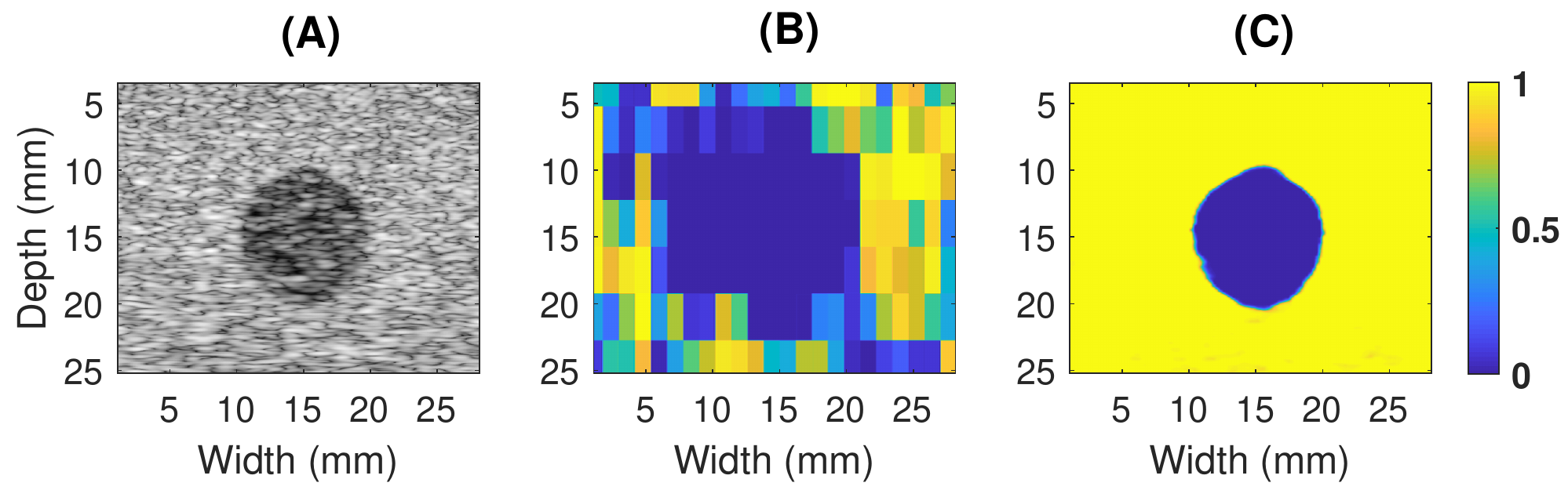}
	\centering
	\caption{The Field II simulation results. B-mode image (A), the segmentation results of the patch-based CNN (B) and our segmentation method (C).}
	\label{fig:field}
\end{figure}

A numerical phantom is simulated using Field II simulation toolbox \cite{jensen1996field}. The transmit focus is at 15 $mm$, the F number is 2.12 and other imaging parameters are the same as \cite{tehrani2021ultrasound}. The proposed method is compared with the patch-based CNN trained on the Field II dataset with the same imaging parameters. In contrast, the proposed method is only trained on the proposed fast simulation method. The aim of this evaluation is to investigate the performance of the segmentation network trained on the simplified fast US simulation on data obtained by the Field II toolbox. The phantom has an inclusion with 2 scatterers per resolution cell and the background having 11 scatterers per resolution cell. The results are illustrated in Fig. \ref{fig:field}. 
The patch-based method cannot clearly identify the boundary of the inclusion, while our proposed method clearly detects the inclusion. Another point is that the patch-based CNN miss-classifies some of the patches outside of the boundary of the inclusion while, the proposed method obtains consistent segmentation of the phantom.

	\subsection{Phantom Results}
	The networks with MTL output (Baseline) and without it (Baseline-nm) are employed for the evaluation. The Baseline is also adapted to the new domain using AdaBN technique (Section \ref{sec:adabn}) to evaluate the performance when BN layers are updated by the new domain (Baseline+AdaBN). Figure \ref{fig:arch} part 3 illustrates the general framework used for experimental phantoms. We also used the recent reference phantom method \cite{Rosado2016} and the patch-based CNN \cite{tehrani2021ultrasound} for comparison (Section \ref{sec:ref}).    
	
	\subsubsection{Homogeneous Phantoms Results}
\begin{figure}[!t]
		\centering
		\includegraphics[width=0.99\textwidth]{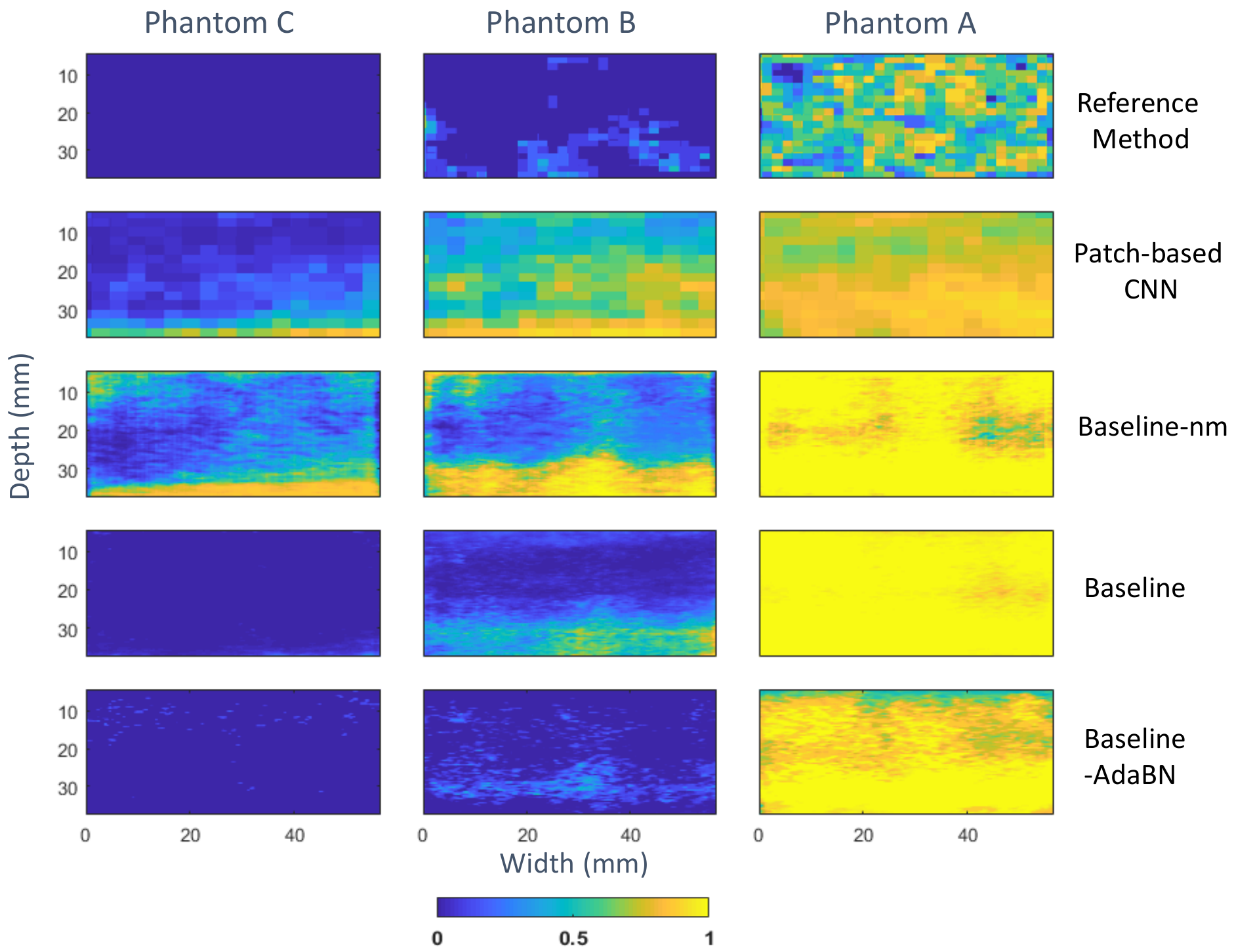}
		\centering
		\caption{Homogeneous phantom results. The color code represents the predicted output of the networks, from 0 (UDS) to 1 (FDS). Correct classes are 0 (UDS) for phantoms C and B, and 1 for phantom A. Using Nakagami parametric image as an axillary output (Baseline) substantially improves the accuracy compared to the network without MTL (Baseline-nm). }
		\label{fig:exp_in}
	\end{figure}

	Baseline, Baseline-nm, Baseline+AdaBN, patch-based CNN (DenseNet + deep supervision) and the reference methods are compared for Phantoms A, B and C. AUC, Accuracy and F1 ($2\frac{precision\times sensitivity}{precision+  sensitivity}$) \cite{fawcett2006introduction} are used as the quantitative metrics and given in Table \ref{tab:phantom}. The average of the results over 8 frames are shown in Fig. \ref{fig:exp_in}. Adding the Nakagami parametric image (MTL) improves the results (compare Baseline and Baseline-nm) which demonstrates that MTL helps networks to be more robust to domain shift. It can be seen that the network trained only on the generated dataset performs well on this dataset without having any domain-specific information which demonstrates the strength of the proposed method. However, the network detected that the bottom of the phantom B to have relatively high probability of FDS. 

	The reference phantom used for this evaluation has the same scatterer number density as the phantom A (high). The imaging setting is also the same. The reference phantom method can discriminate well phantoms C and B. Also, it performs acceptable on phantom A. It should be mentioned that this method requires patching and each patch needs to be compared with reference phantom patch from the same depth to consider its variations. Our network with adaBN (Baseline+AdaBN) performs better than Baseline especially for phantom B. Furthermore, it outperforms the reference method on phantom A. It should be noted that two frames of phantom A and B are utilized for updating the BN layer statistics and weights of network are kept fixed. The BN impact can be observed by comparing the Baseline and Basline+AdaBN. The only difference between these two networks is that in the latter BN statistics are updated by the test domain.
	
	The reference method requires a reference phantom imaged by the same ultrasound machine which may be not available. However, the patch-based CNN and our proposed network trained on generated dataset (Baseline) performs well without any information about the test domain. Comparing the proposed method and the patch-based CNN, Baseline performs substantially better than the patch-based CNN especially on the phantom B which is more challenging than phantom C and A. This demonstrates that the proposed method is more robust to the change of domains compared to the patch-based CNN due to the fact that the network is trained using the training dataset with diverse imaging parameters.

	\begin{table}[]
		\caption{Homogeneous phantom results}
		\label{tab:phantom}
		\resizebox{0.8\textwidth}{!}{%
			\begin{tabular}{@{}cccc@{}}
				\toprule
				& \textbf{AUC} & \textbf{Accuracy} & \textbf{F1} \\ \midrule
				Reference Method & 0.987        & 0.912             & 0.848       \\
				Patch-based CNN  & 0.780 & 0.740 & 0.678 \\
				Baseline-nm          & 0.974        & 0.797             & 0.766         \\
				Baseline          & 1.00        & 0.957             & 0.934         \\
				Baseline+AdaBN    & 1.00         & 0.999             & 0.998       \\ \bottomrule
			\end{tabular}%
		}
	\end{table}
	
	\begin{figure}
		
		\centering
		\begin{subfigure}{0.32\linewidth}
			
			\centering
			\includegraphics[width=0.11\textheight]{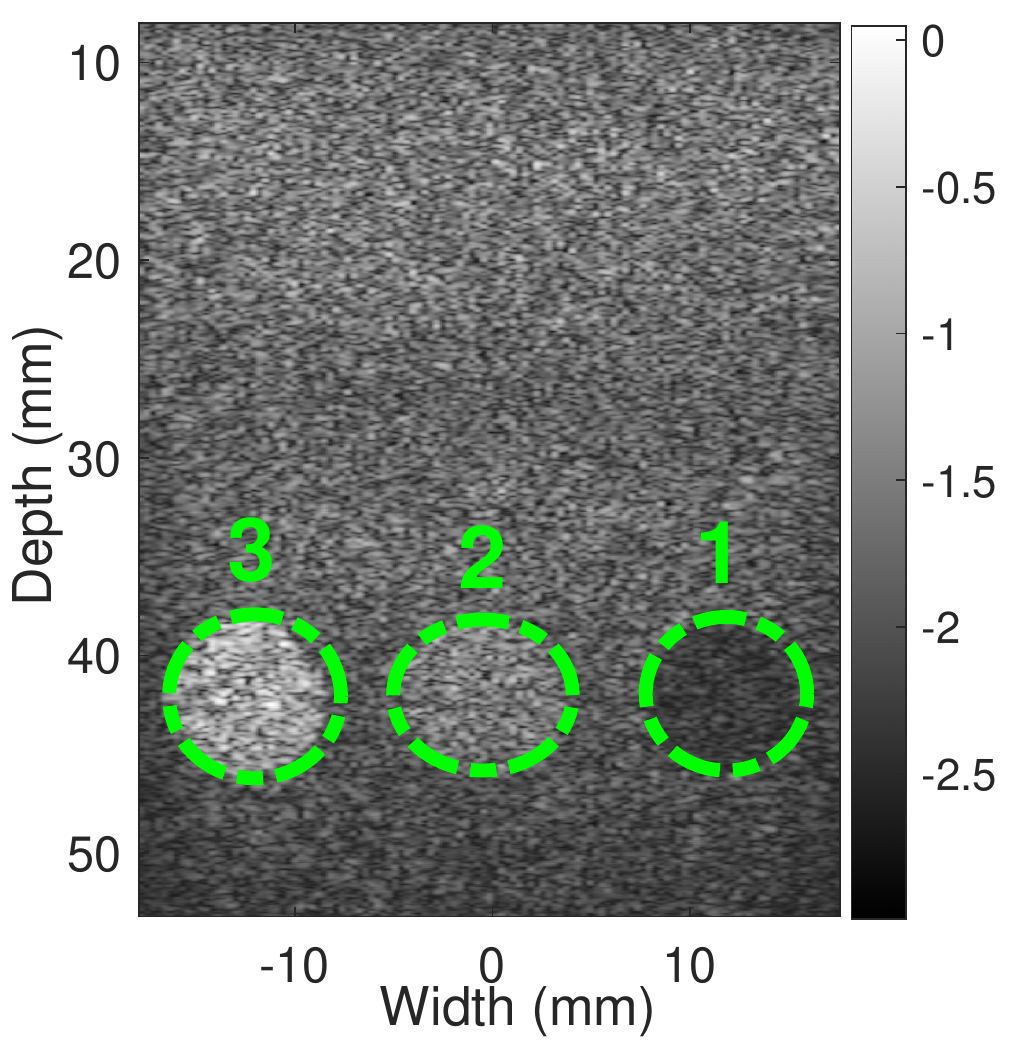}
			\caption{}
			\label{fig:phantomd_bmode}
		\end{subfigure}
		\begin{subfigure}{0.32\linewidth}
			\centering
			\includegraphics[width=0.11\textheight]{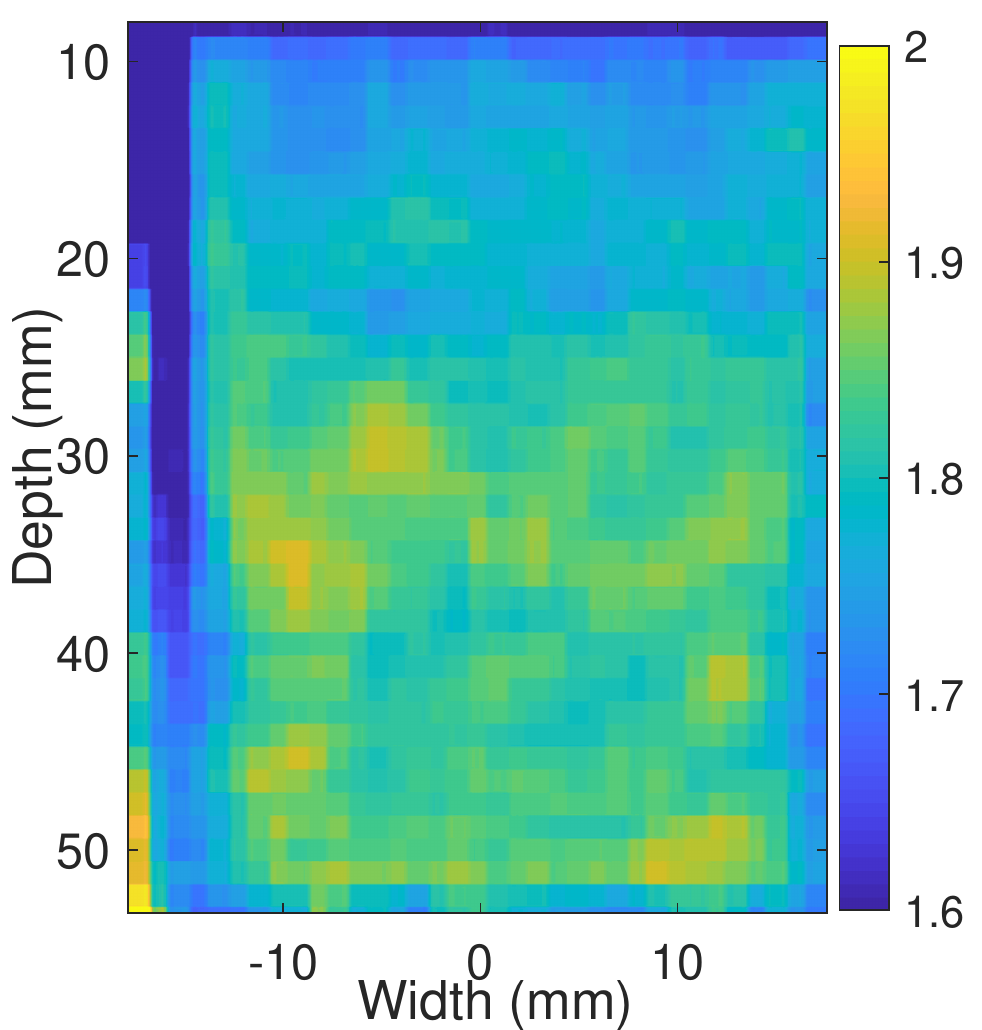}
			\caption{}
		\end{subfigure}
		\begin{subfigure}{0.32\linewidth}
			\centering
			\includegraphics[width=0.11\textheight]{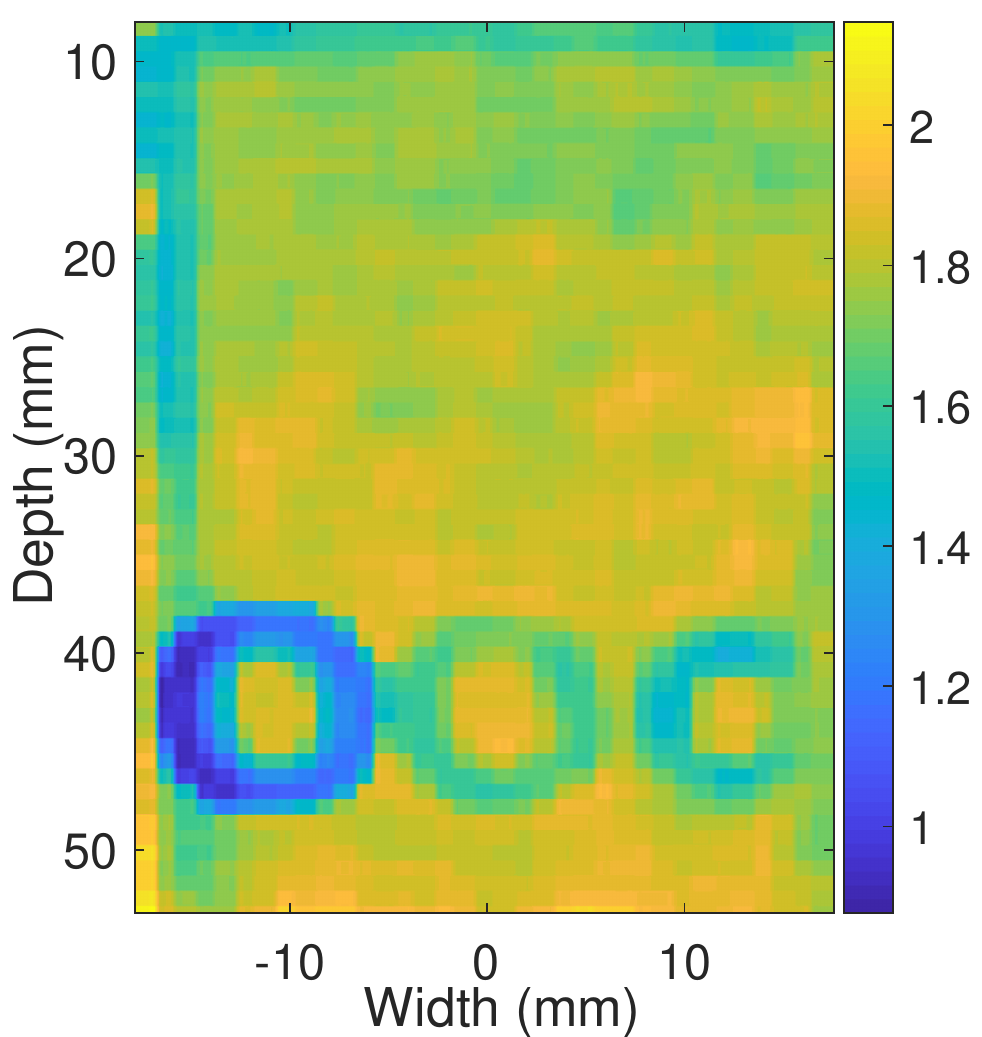}
			\caption{}
		\end{subfigure}
		\begin{subfigure}{0.32\linewidth}
			\centering
			\includegraphics[width=0.11\textheight]{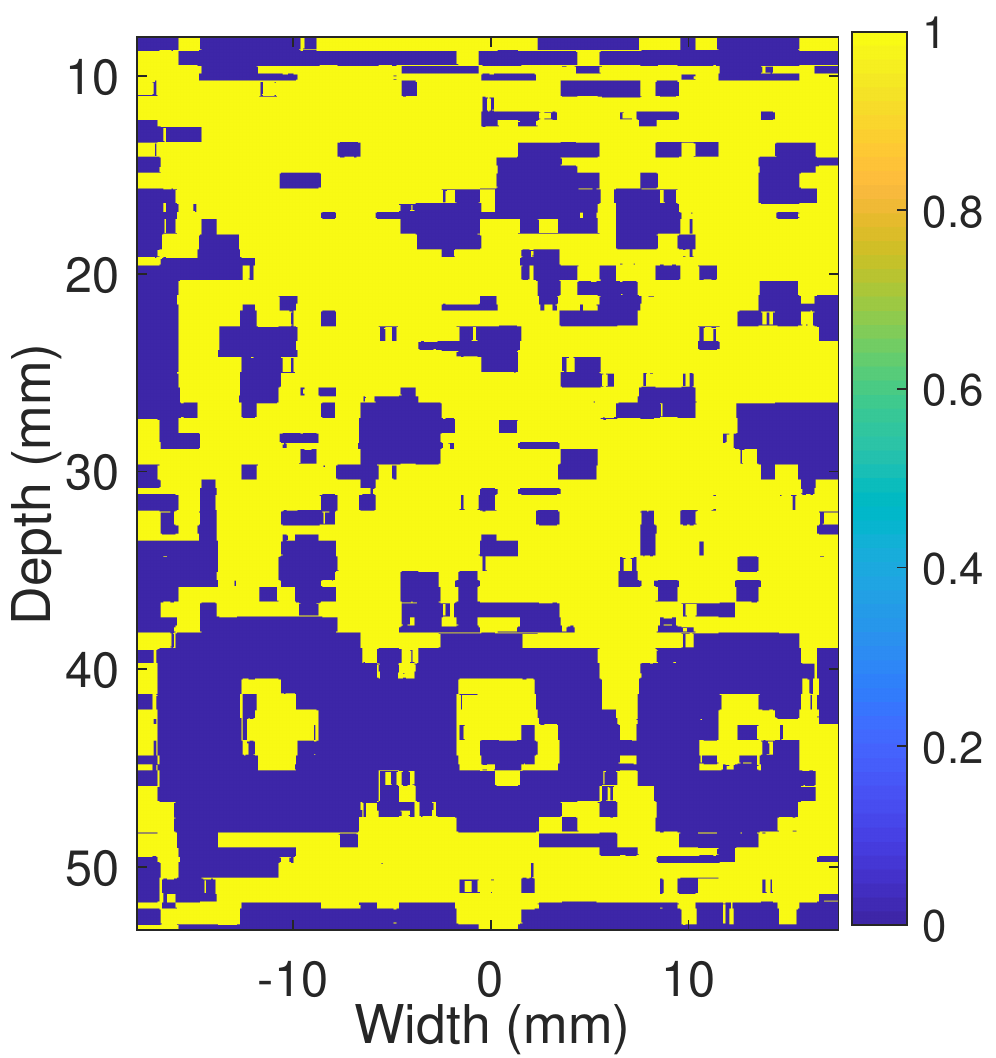}
			\caption{}
		\end{subfigure}
		\begin{subfigure}{0.32\linewidth}
			\centering
			\includegraphics[width=0.11\textheight]{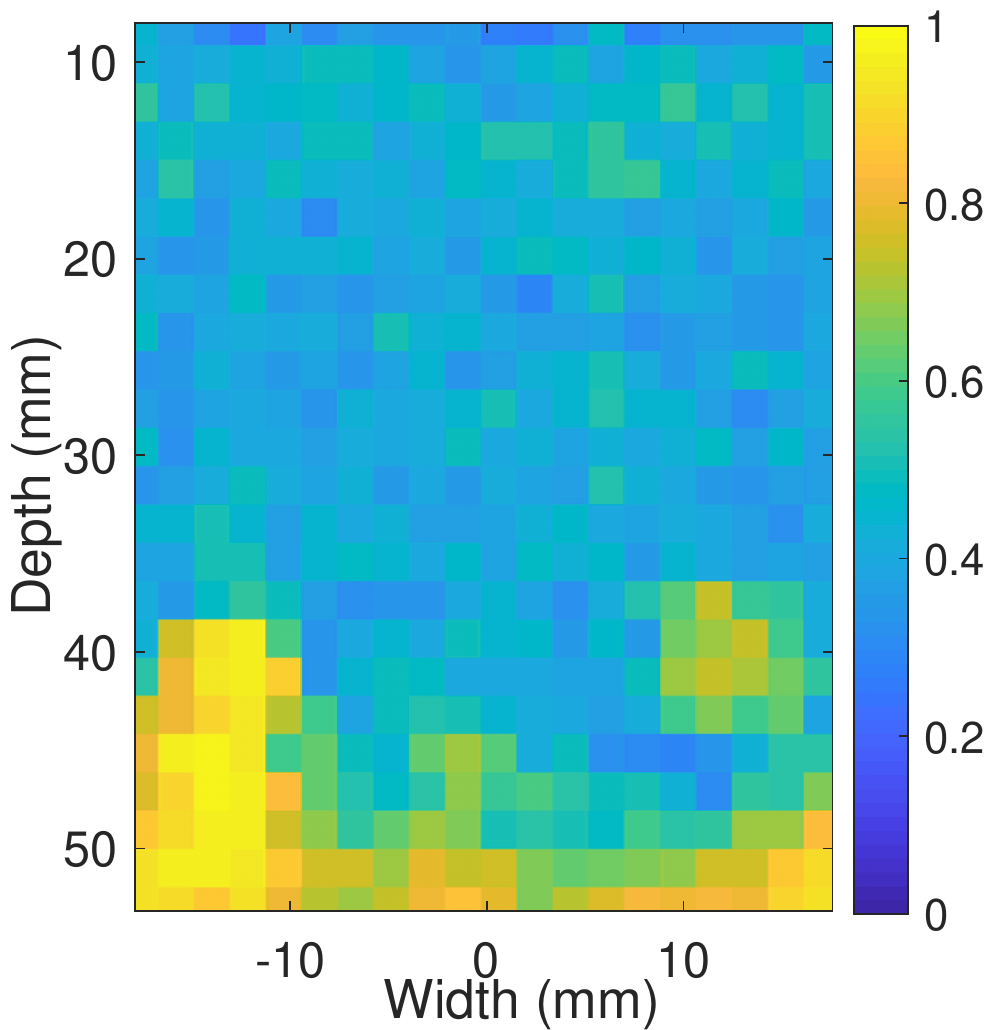}
			\caption{}
		\end{subfigure}
		\begin{subfigure}{0.32\linewidth}
			\centering
			\includegraphics[width=0.11\textheight]{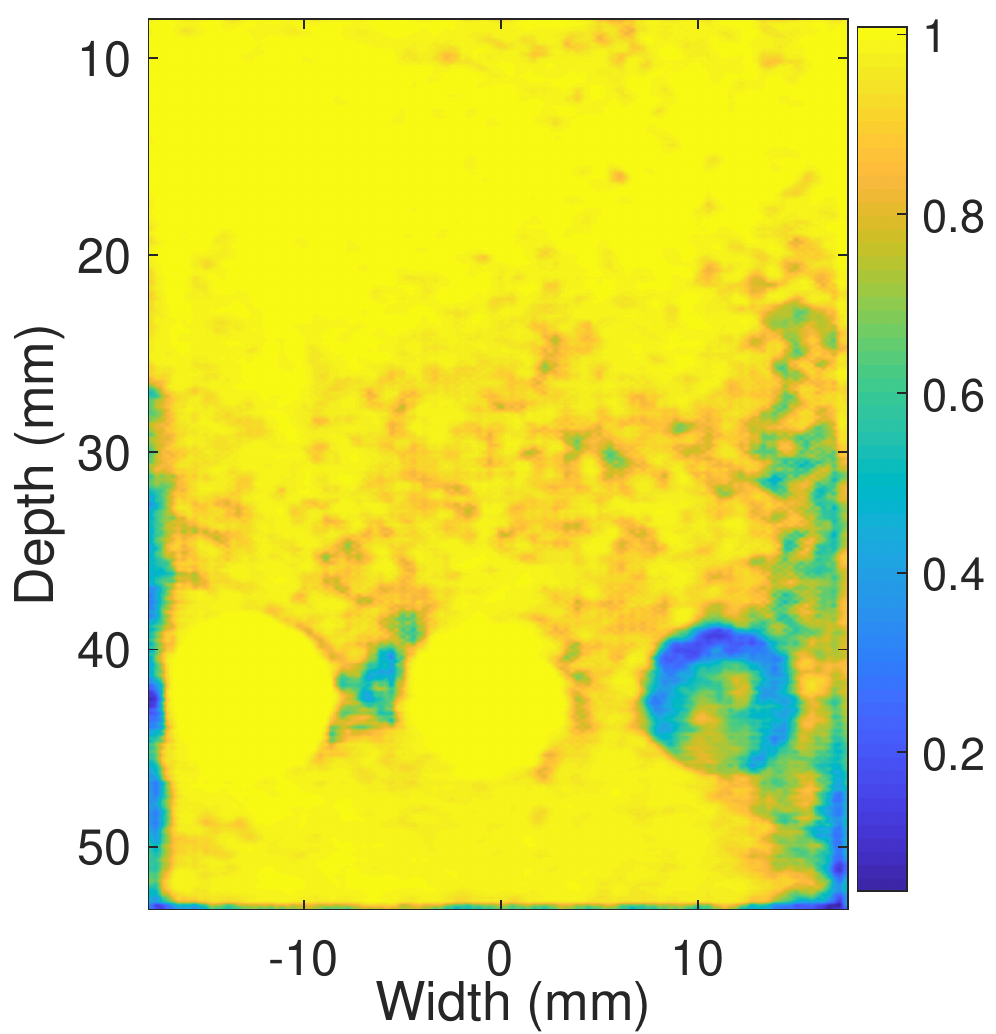}
			\caption{}
		\end{subfigure}

		\vspace{-0.01cm}
		\caption{The phantom D results. B-mode image (a), SNR (mean) of reference phantoms (b), SNR parametric image (c), reference phantom method (d), patch-based CNN (e) and the proposed method (Baseline) (f). Inclusions are specified in the B-mode image. Inclusion 1 belongs to UDS class and other parts belong to FDS class. The inclusions are numbered from the lowest to the highest scatterer number density. }
		\label{fig:phantomd}
	\end{figure}  
	\subsubsection{Phantom D Results}
	The B-mode image of the phantom with inclusions is shown in Fig. \ref{fig:phantomd_bmode}. This phantom contains three inclusions which are specified in the figure. The scattering agent of inclusions and the background are the same (mean scatterer amplitude is fixed); therefore, the intensity can be associated to the density of scatterers. We know \textit{a priori} that the background is FDS. The inclusions 2 and 3 have higher intensity which means that the have higher scatterer number density than background; therefore, they are also FDS. The inclusion 1 has lower density and it is UDS. For the reference method, we use other part of the phantom without inclusion having high scatterer number density imaged by the same machine as the reference phantom. The SNR parametric image of reference phantom is obtained by averaging the SNR parametric image of 10 frames. The SNR of the reference phantom is illustrated in Fig. \ref{fig:phantomd} (b). We also obtained the average SNR of 12 frames of the phantom D which are depicted in Fig. \ref{fig:phantomd} (c). The SNR value of the reference phantom is 1.725$\pm$0.225 which is expected for phantoms with high scatterer number density. However, SNR of phantom D is as low as 1 in some regions. The main reason is that SNR on the borders of the regions with different scatterer number densities, is not valid and reliable since on the borders of the inclusions the patch contains two different distribution and the calculated value is not reliable anymore. Therefore, patch-based methods usually fail in these regions. The output of the reference phantom method is shown in Fig. \ref{fig:phantomd} (d). As anticipated, most of the background are correctly predicted as FDS. While, the inclusions 2 and 3 borders are incorrectly classified as UDS. It should be noted that the drop of the predicted scatterer number density between the inclusions 2 and 3 is an artifact.  It can be caused by side lobes, off-axis scattering and having limited number of samples between the two inclusion. It is worth mentioning that in the grid-based simulation, the echogenicity is directly determined by the tissue scattering (mean scattering and the scatterer number density). Therefore, artifacts such as refraction effects caused by rounded boundaries and shadowing and enhancement caused by attenuation could provide unreliable classifications.

	The Baseline method which is trained on simulation data, is employed for obtaining the predicted scatterer number density mask. The output segmentation mask of the proposed method average and standard deviation across 12 frames are shown in Fig. \ref{fig:phantomd} (e) and (f). The patch-based methods correctly shows that the inclusion 3 has higher scatterer number density than the other regions. However, it fails to detect the lower scatterer number density of inclusion 1. Our proposed network correctly classifies the background and inclusions 2 and 3. Inclusion 1 can be also well discriminated from the other two inclusions. It should be mentioned that no reference phantoms have been employed for the proposed method. Without any reference phantom requirement, the proposed method outperforms the conventional reference phantom method which shows the potential of CNNs in QUS analysis.
	
	\subsection{Phantom E Results}

	\begin{figure}[!t]
		\centering
		\includegraphics[width=0.99\textwidth]{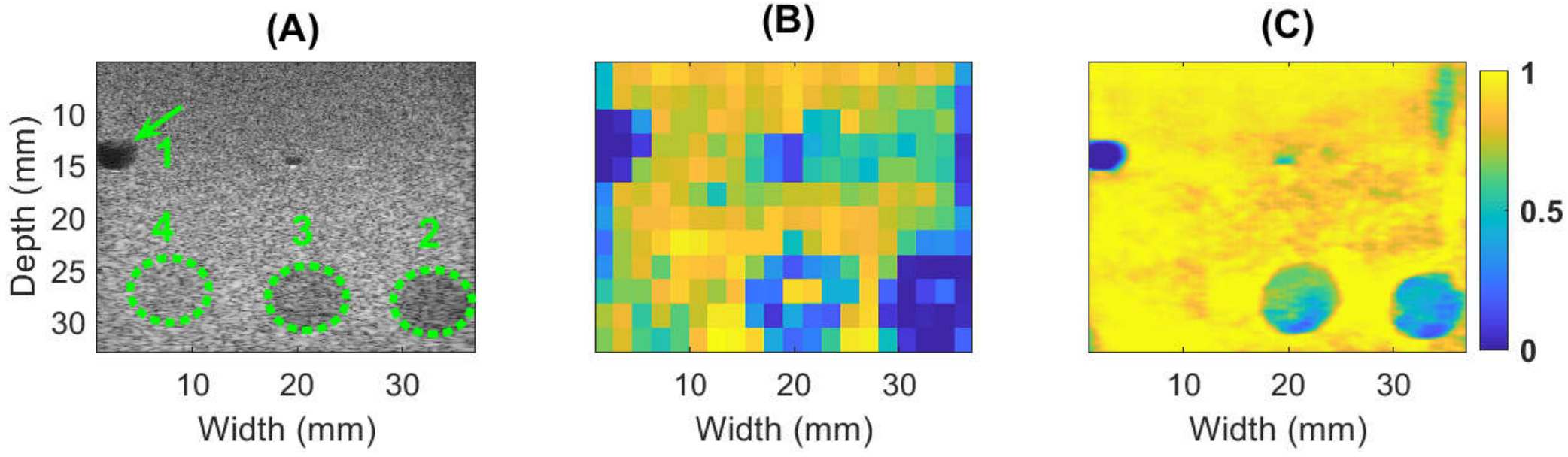}
		\centering
		\caption{Scatterer-density segmentation of the CIRS phantoms scanned with the Alpinion system. (A) B-mode image, (B) segmentation result using the patch-based CNN \cite{tehrani2021ultrasound}, and (C) segmentation using the patch-less CNN (Baseline). Inclusions are numbered from the lowest to the highest scatterer number density. The color bar indicates probability of FDS. }
		\label{fig:alp1}
	\end{figure} 	
	
The average of the results over 25 frames are shown in Fig. \ref{fig:alp1}. As shown in the B-mode image, there are 4 inclusions. The first inclusion is anechoic, while the other inclusions have fewer scatterer number density compared to the background. The inclusions are numbered from the lowest scatterer number density to the highest. The patch-based CNN can detect the approximate location of inclusions 1, 2 and 3. However, the boundaries of them are not accurately identified due to the finite size of the patch. The proposed method provides substantially higher quality scatterer number density segmentation compared to the patch-based CNN. More experiments for this phantom using different scanners are available in the Supplementary Materials, which show the robustness of the proposed method.

It should be mentioned that the network is also able to discriminate the inclusions 1 and 2 in Fig. \ref{fig:alp1}. The average of the network's output for the inclusion 1 and 2 are 0.441 and 0.577, respectively, which indicates that the network gives higher probability of FDS for the inclusion that has the higher  scatterer  number density. This can be employed to characterize the US images having scatterer number densities between 0-10 within the resolution cell.

	\subsection{\textit{In vivo} Results}
\begin{figure}[!t]
		\centering
		\includegraphics[width=0.95\textwidth]{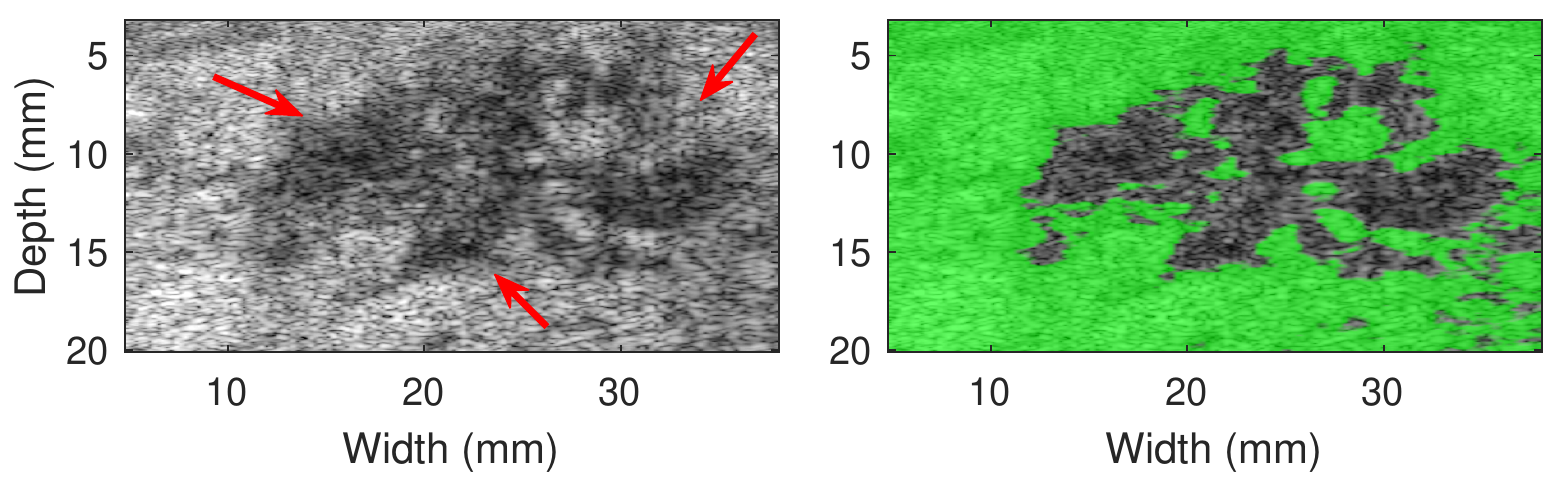}
		\centering
		\caption{Scatterer-density segmentation of \textit{in vivo} breast fibroadenoma (1). B-mode image (left) and the segmentation results of the proposed method overlaid on the B-mode image (right). The green mask denotes predicted FDS regions. The fibroadenoma tumor is specified on the B-mode image. }
		\label{fig:invivo1}
	\end{figure}

\begin{figure}[!t]
		\centering
		\includegraphics[width=0.87\textwidth]{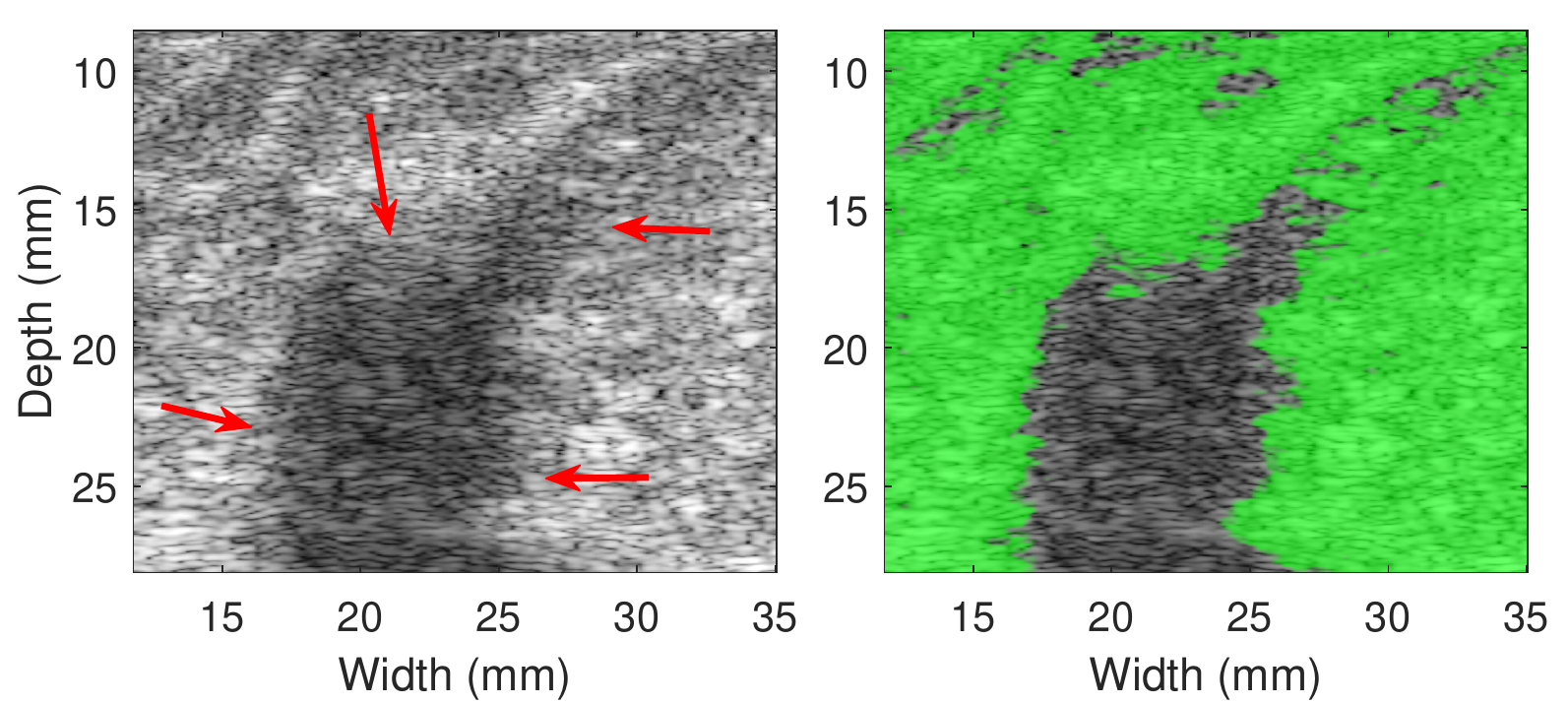}
		\centering
		\caption{Scatterer-density segmentation of \textit{in vivo} breast fibroadenoma (2). B-mode image (left) and the segmentation results of the proposed method overlaid on the B-mode image (right). The green mask denotes predicted FDS regions. The fibroadenoma tumor is specified on the B-mode image.}
		\label{fig:invivo2}
	\end{figure}

The patch-less CNN segmentation was tested on echo signals acquired from two breast fibroadenomas. Reference phantoms with known scatterer number density imaged using the same machine setting are not available for this dataset. According to \cite{tsui2008classification}, the Nakagami \textit{m} parameter, which is highly correlated with scatterer number density, of fibroadenomas tends to be lower than values in normal breast tissues, which have high scatterer number density close to FDS limit (here, we only considered fat regions and other regions such as ductal cells are excluded from our study). Therefore, the network should be able to discriminate fibroadenomas and normal breast tissue. The results are presented in Figs. \ref{fig:invivo1} and \ref{fig:invivo2}. The network classifies normal breast tissue as FDS; while, the regions of the fibroadenomas are segmented as UDS. The patch-based CNN method was also used in this data but it failed at detecting the lesion from the background. These results are included in Supplementary Materials (Fig. S4 and S5).
	
It should be noted that knowing the exact ground truth scatterer number density of the \textit{in vivo} data requires acoustic microscopy image analysis which is out of scope of this work.     
	
	\section{Discussion}
	Segmentation of the density of scatterers is very important for estimation of other QUS parameters since different considerations must be taken for areas with low scatterer number density \cite{tehrani2021ultrasound}. In this paper, the scatterer number density of US images are segmented. The mean scatterer amplitude and scatterer number density both contribute to the intensity of envelope data. Therefore, intensity cannot be used as a reliable feature and both must be taken into account during the training. 
	
	In this paper, we generated a very large dataset having varied mean scatterer amplitude and scatterer number density. To the best of our knowledge, this is the first time such a large dataset is generated which considers that both mean scatterer amplitude and scatterer number density are varied. The generated dataset can be used to train different networks. The imaging settings are also varied and randomly sampled from  uniform distributions to increase the generalization of the dataset and improve the robustness of the network when images are obtained from different machine settings. Generating such a large dataset (15000 images) is computationally prohibitive with ultrasound simulation toolboxes such as Field II. Therefore, a grid-based method was employed to generate such a large dataset within manageable times. Field II takes 50 minutes to simulate a phantom (with imaging settings of \cite{tehrani2021ultrasound}), whereas our simplified grid-based method only takes 1 second to simulate the same phantom. 
	
	The performance of the network without having any information about the test domain was investigated. The Nakagami parametric image was employed as an additional output to reduce the system dependency. The network trained by the generated dataset was able to correctly classify the homogeneous phantoms and segment the phantom with inclusions without any information about the test domain.
	
	The statistical parameters used in quantifying scatterer number density are system dependent. Reference phantoms have been utilized to reduce the effects of machine settings. We investigated how to use a few frames of test domain to update the network. We showed that updating BN layers statistics is enough to adapt the network to the new domain. The statistics of BN layers was updated using a reference set which is a mini-batch composed of data belonging to both classes. Unlike transfer learning, this method only needs a few frames of both classes which facilitates the utilization of reference phantoms for CNNs. The network adopted by the reference set (Baseline+AdaBN) performs better than the network trained on generated dataset especially for the phantom B.  
	
	Considering phantom D, We evaluated the reference phantom method using SNR of a reference phantom averaged across 10 frames as well as a recent patch-based CNN. The previously proposed reference phantom method was able to correctly classify most of background regions. However, the inclusions especially on the borders were not classified correctly. Also, the patch-based CNN was able to detect inclusion 3 but fails on the other inclusions. Also, it did not predict the background as FDS with a high confidence (the average of the output of the network for background region was around 0.5). The reason was that the network was trained on fixed imaging settings. The proposed method correctly segmented this phantom without using any patching. Our results confirmed that the proposed segmentation method was more robust than the patch-based CNN since it was trained on a large dataset comprised of diverse imaging settings, scatterer number densities and mean scattering amplitudes.  
	
	The recent patch-based CNN \cite{tehrani2021ultrasound} is also employed for evaluation. The AUC of this method for homogeneous phantom results (0.78) was lower than the value obtained in \cite{tehrani2021ultrasound}. The reason is that in \cite{tehrani2021ultrasound}, the results are reported for single frame while, in this paper the average of the results over multiple frames are reported. The drop of performance of patch-based method can also be observed in Fig. \ref{fig:field} due to the beam divergence in the bottom of the simulated phantom.  
	
	This work demonstrates the potential of using CNNs for QUS. The goal of the paper is not evaluating different available segmentation networks for scatterer number density segmentation but to provide a general framework to segment scatterer number density using CNNs. Other segmentation networks can also be employed for this task. To the best of our knowledge, this is the first work that segments the scatterer number density of the whole image without patching. 
	
	The proposed method has various limitations. First, the simulation approach used to generate the training dataset assumed a homogeneous point spread function within each image. However, different PSF sizes were used in generating different images in the training set. In practice, the PSF changes with depth, resulting in a variation of the effective number of scatterers inside the resolution cell. Despite this limitation, the simplified model shows promising results in Field II simulations and experimental phantoms where the PSF changes because the task here is the segmentation based on FDS and UDS, and not estimating the scatterer number density per resolution cell.            
	
	\section{Conclusion}
	In this paper, the scatterer number density of ultrasound images was segmented using CNN without any patching. To be able to train the network, a large and diverse dataset was required hence different shapes of mean scatterer amplitude and scatterer number density was considered to generate this dataset. We also investigated how to use statistical parametric images such as the Nakagami parametric image to improve the performance in presence of domain shift. In addition to this, inspired by reference methods used in QUS algorithms, we proposed to use a reference set composed of a few frames from both classes. This reference set was utilized to update the statistics of BN layers. We showed that updating the BN layers is adequate and there is no need to update the network weights (fine-tuning) which substantially reduces the amount of data required for domain adaptation. The method was tested on five experimental phantoms and two \textit{in vivo} data imaged by different US scanners. The network was only trained on simulations using the grid approach. Despite this limitation, our method was able to segment the scatterer number density of different simulated data, experimental phantoms and \textit{in vivo} data.   
	\vspace{-0.01cm}
	\section*{Acknowledgment}

	\label{sec:acknowledgments}
	We thank Dr. Timothy J. Hall for invaluable discussions. We acknowledge the support of the Natural Sciences and Engineering Research Council of Canada (NSERC) RGPIN-2020-04612, NIH R01HD072077, and UNAM PAPIIT IA102320.

	\bibliographystyle{IEEEtran}
	
	\bibliography{refs3}

\end{document}


\title{Supplementary Material for Robust Scatterer Number Density Segmentation of Ultrasound Images}
\author{Ali K. Z. Tehrani, Ivan M. Rosado-Mendez, and Hassan Rivaz
\thanks{This work is supported by Natural Sciences and Engineering Research Council of Canada (NSERC) RGPIN-2020-04612, NIH R01HD072077, and UNAM PAPIIT IA102320.}

\thanks{ }
\thanks{A. K. Z. Tehrani and H. Rivaz are with the Department
	of Electrical and Computer Engineering, Concordia University, Canada.	
	Ivan M. Rosado-Mendez is with the Department of Medical Physics, University of Wisconsin, United States.
	e-mail: A\_Kafaei@encs.concordia.ca, rosadomendez@wisc.edu, and 
	hrivaz@ece.concordia.ca}}

\maketitle

\section{More results for phantom E}
\begin{figure*}[!t]
	\centering
	\renewcommand\thefigure{S1}
	\includegraphics[width=0.85\textwidth]{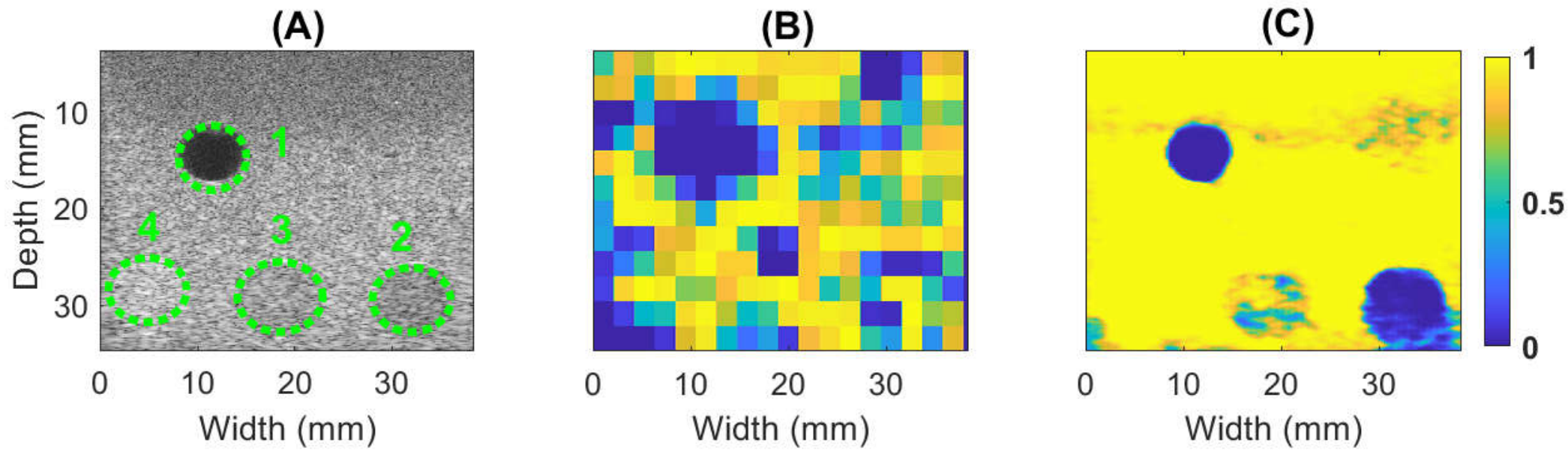}
	\centering
	\caption{The result of phantom E using Alpinion scanner. B-mode image (A), the segmentation results of the patch-based CNN (B) and our patch-less segmentation method (C). }
	\label{fig:alp2}
\end{figure*}

\begin{figure*}[t]
		\renewcommand\thefigure{S2}
	\centering
	\includegraphics[width=0.85\textwidth]{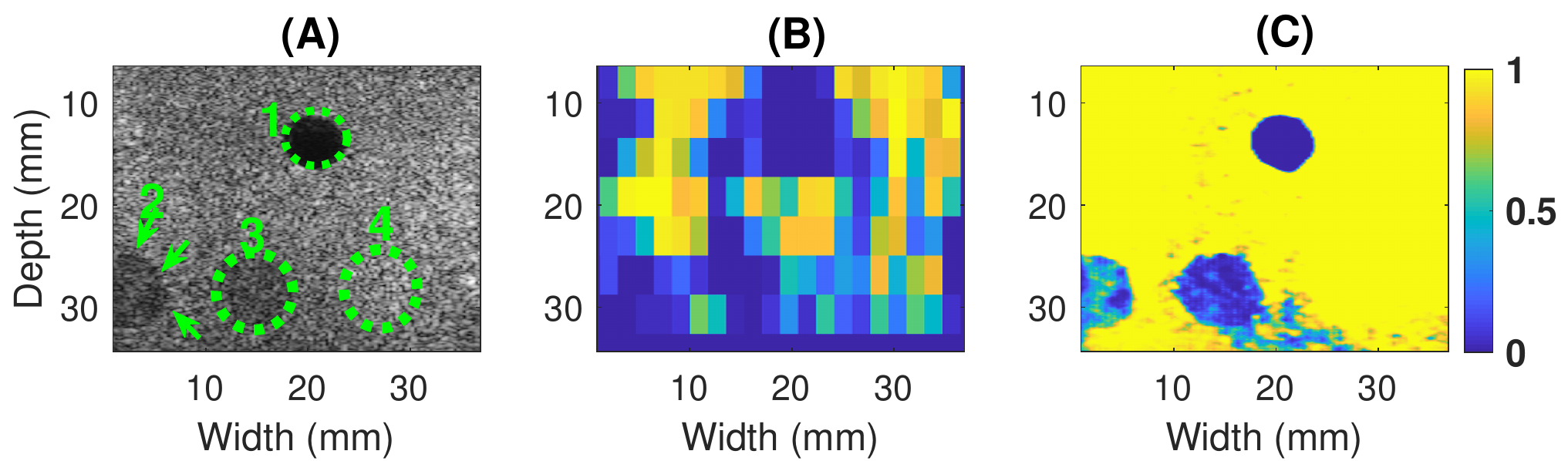}
	\centering
	\caption{The result of phantom E using Verasonic scanner (1). B-mode image (A), the segmentation results of the patch-based CNN (B) and our patch-less segmentation method (C).}
	\label{fig:verasonic}
\end{figure*}

\begin{figure*}[t]
		\renewcommand\thefigure{S3}
	\centering
	\includegraphics[width=0.85\textwidth]{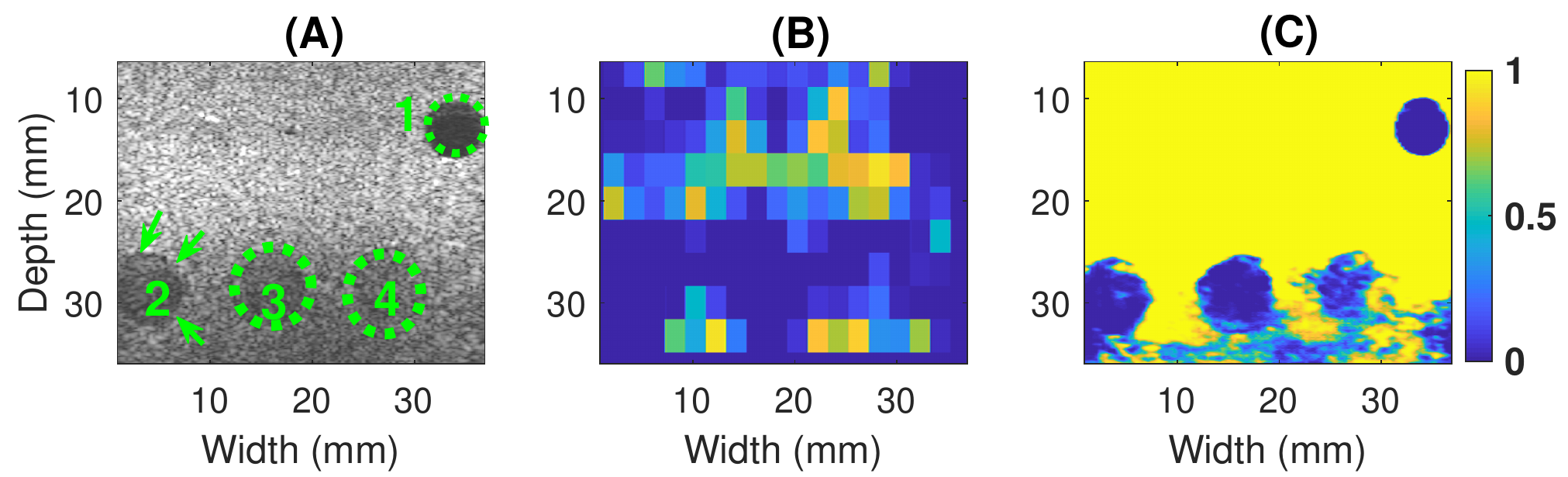}
	\centering
	\caption{The result of phantom E using Verasonic scanner (2). B-mode image (A), the segmentation results of the patch-based CNN (B) and our patch-less segmentation method (C).}
	\label{fig:verasonic_2}
\end{figure*}

The results of data collected from phantom E using the E-CUBE 12 Alpinion machine are given in the paper. Here, more data is collected from other parts of this phantom using E-CUBE 12 Alpinion machine Fig. \ref{fig:alp2} and Verasonic Vantage 256 Fig. \ref{fig:verasonic} and \ref{fig:verasonic_2}. The results shows that the patch-based CNN is unable to detect the inclusions having low scatterer number density clearly whereas, the proposed method provides high-quality segmentation of the inclusions and the background.

 It should be mentioned that the network detects the inclusion 3 in Fig. \ref{fig:alp2} but not as clearly as the other ones since it has close scatterer number density to the background. The inclusion 3 in Fig. \ref{fig:alp2} can be discriminated from the inclusion 1 and 2 as well as the background since the network outputs higher values for this inclusion (average of the network output for this inclusion is 0.714) than the inclusions 1 and 2 (average of the network output for inclusions 1 and 2 are 0 and 0.03, respectively) and lower than the background.            

\section{\textit{In vivo} Results of Patch-based CNN}
\begin{figure}[!t]
 			\renewcommand\thefigure{S4}
 	\centering
 	\includegraphics[width=0.95\textwidth]{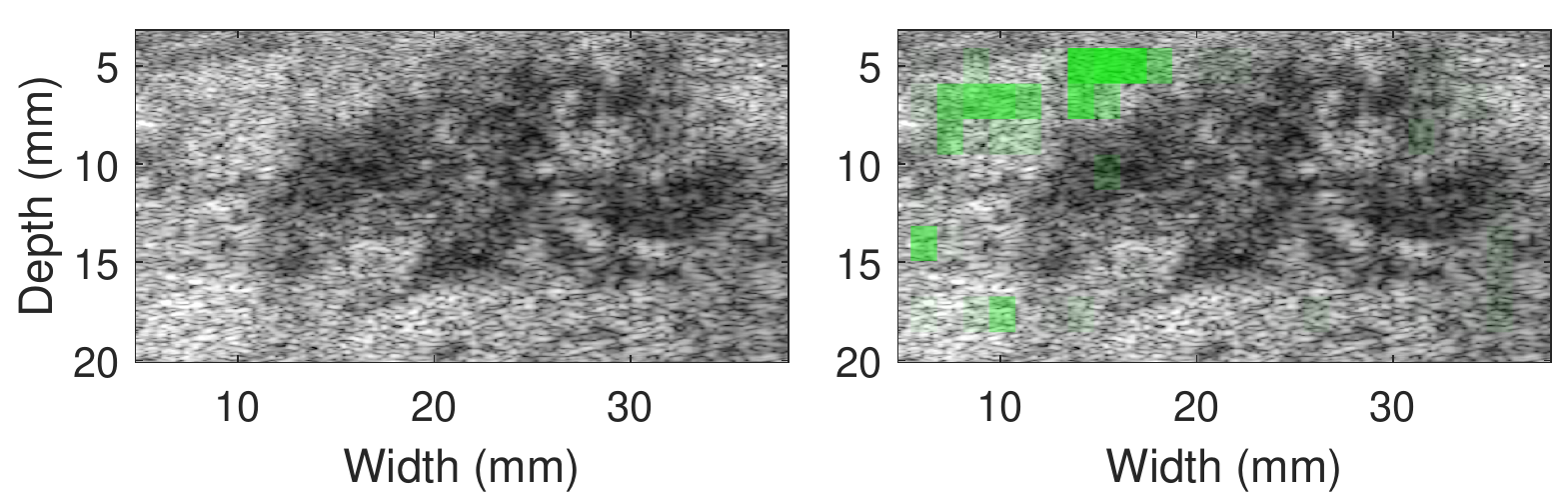}
 	\centering
 	\caption{Scatterer-density segmentation of \textit{in vivo} data 1. B-mode image (left) and the segmentation results of the patch-based CNN overlaid on the B-mode image (right). The green mask denotes predicted FDS regions.}
 	\label{fig:invivo1}
 \end{figure}

\begin{figure}[!t]
 	\renewcommand\thefigure{S5}
 	\centering
 	\includegraphics[width=0.87\textwidth]{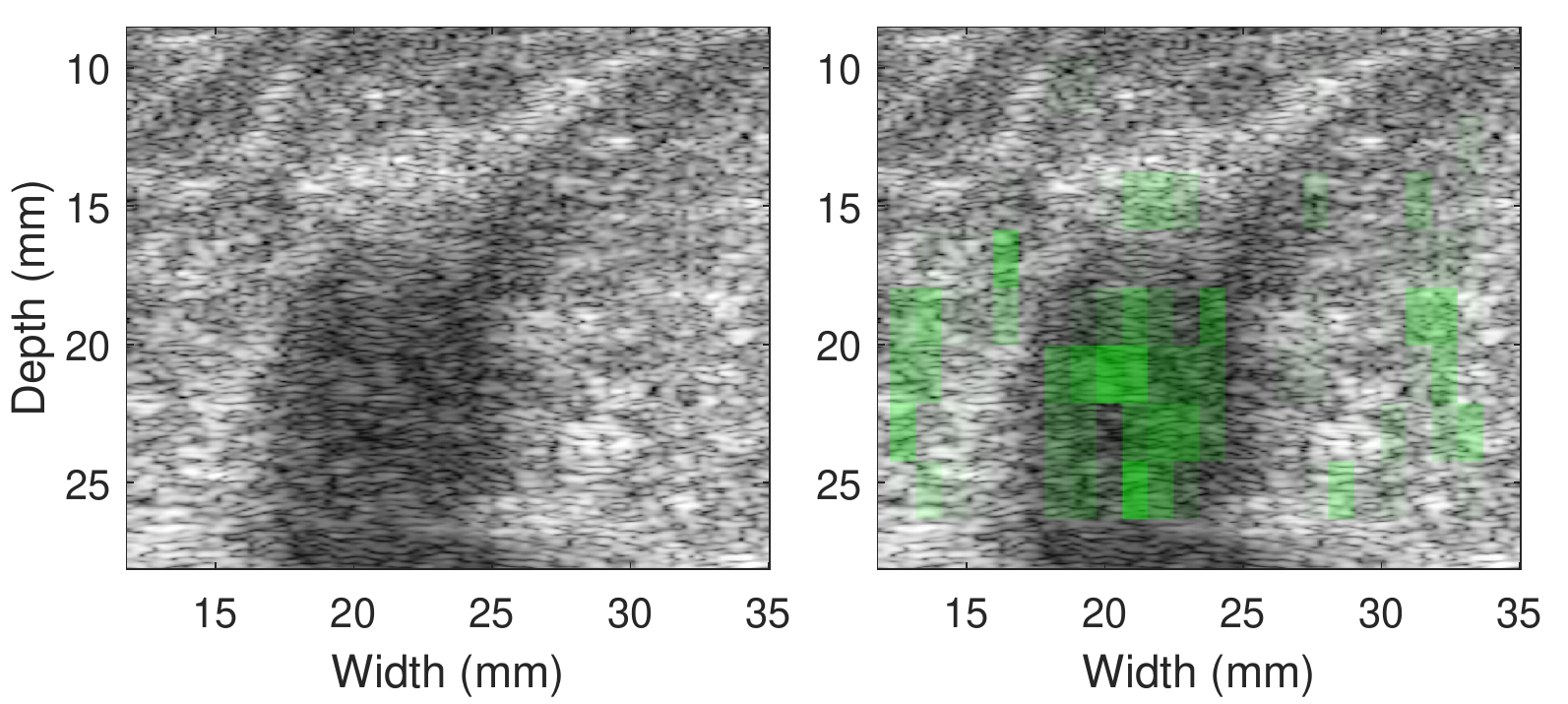}
 	\centering
 	\caption{Scatterer-density segmentation of \textit{in vivo} data 2. B-mode image (left) and the segmentation results of the patch-based CNN overlaid on the B-mode image (right). The green mask denotes predicted FDS regions.}
 	\label{fig:invivo2}
 \end{figure}

The results of the patch-based CNN are illustrated in Fig. \ref{fig:invivo1} and \ref{fig:invivo2}. The patch-based CNN fails to detect the high scatterer number density of the breast masses. It detects most regions as UDS. The networks fails due to the high variability scattering properties of the \textit{in vivo} data and the large domain shift presented between training and the \textit{in vivo} data.




